\documentclass[aps,pre,groupedaddress,twocolumn,nofootinbib]{revtex4-1}
\usepackage[pdftex]{graphicx}
\usepackage{hyperref}

\begin{document}



\title{Branching Dynamics of Viral Information Spreading}


\author{Jos\'e Luis Iribarren}
\email[]{jose.iribarren@iic.uam.es}
\affiliation{Instituto de Ingenier\'ia del Conocimiento, Universidad Aut\'onoma de Madrid, 28049 Madrid, Spain}

\author{Esteban Moro}
\email[]{emoro@math.uc3m.es}
\affiliation{Departamento de Matem\'aticas \& GISC, Universidad Carlos III de Madrid, 28911 Legan\'es (Madrid) \\ Instituto de Ciencias Matem\'aticas CSIC-UAM-UCM-UC3M, 28049 Madrid, Spain\\Instituto de Ingenier\'ia del Conocimiento, Universidad Aut\'onoma de Madrid, 28049 Madrid, Spain}

\date{\today}

\begin{abstract}

Despite its importance for rumors or innovations propagation, peer-to-peer collaboration, social networking or Marketing, the dynamics of information spreading is not well understood. Since the diffusion depends on the heterogeneous patterns of human behavior and is driven by the participants' decisions, its propagation dynamics shows surprising properties not explained by traditional epidemic or contagion models. Here we present a detailed analysis of our study of real Viral Marketing campaigns where tracking the propagation of a controlled message allowed us to analyze the structure and dynamics of a diffusion graph involving over 31,000 individuals. We found that information spreading displays a non-Markovian branching dynamics that can be modeled by a two-step Bellman-Harris Branching Process that generalizes the static models known in the literature and incorporates the high variability of human behavior. It explains accurately all the features of information propagation under the ``tipping-point'' and can be used for prediction and management of viral information spreading processes.

\end{abstract}

\pacs{89.75.-k, 05.10.-a}

\maketitle

\section{Introduction}
Each day, millions of conversations, emails, SMS, blog posts and comments, instant messages, tweets or web pages containing various types of information are exchanged between people. Humans natural inclination to share information with others in a ``viral'' fashion stems from the need of socializing and seeks to gain reputation, influence, trustworthiness or popularity \cite{bruyn}. Such viral dissemination of information through social networks, commonly known as ``Word-of-Mouth'' (WOM), is of paramount importance in our everyday life. In fact, it is known to influence purchasing decisions to the extent that 2/3 of the United States economy is driven by those kind of personal recommendations \cite{buzzbuzz}. WOM is also important to understand sales and customer value \cite{Kumar2007,Schmitt}, opinion formation or rumor spreading in social networks \cite{barrat,castellano} or to determine the influence of each person in its social neighborhood \cite{wattsJCR,influence}. Despite its importance and due to the difficulty (or inability) to capture this phenomenon, detailed empirical data on how humans disseminate information are scarce \cite{iribarren2009}, population aggregated \cite{aggregated} or indirect \cite{indirect1,liben}. Moreover, most studies have concentrated on asymptotical stationary properties of information difussion \cite{epidemic,golub,dwang,cebrian}. This has hampered the study of the dynamics of information diffusion and indeed most of its understanding comes from theoretical propagation models running on empirical or synthetic social networks in an approach borrowed from epidemiology \cite{daley,yamir,motion}. In those models, information diffusion equates to the propagation of virus or diseases that spontaneously pass to others by contagion through the active social connections of the infected (i.e. informed) agents.

\smallskip

However, information diffusion mechanisms are fundamentally different from those operating in disease spreading. In fact, passing a message along has a perceived transmission cost, its targets are consciously selected among potentially interested individuals \cite{huberman,flow}, depends on human volition and, ultimately, is executed on the individuals' activity schedule. An obvious implication of those peculiarities is that information spreading is bound to depend on the large variability observed both on the volume and frequency of human activities and on the perceived value/cost of transmitting the information. For example, the number of emails sent by individuals per day \cite{barabasinature}, the number of telephone calls placed by users \cite{telephone}, the number of blog entries by user \cite{blogs,blogs1}, the number of web page clicks per user \cite{pitkow}, and the number of a person's social relationships \cite{tipping} or sexual contacts \cite{sexual} show large demographic stochasticity. In fact these numbers are distributed according to a power-law (or Pareto) distribution, inconsistent with the mild Gaussian or Poissonian stochasticity around population-averaged values traditionally assumed in epidemiological models \cite{andersonmay}. The same large variability pattern applies to the human activities time dynamics: for example, email response delays, market trading frequencies or inter-event time of web page visits, telephone calls, etc. are well described by power-law or log-normal distributions \cite{barabasinature,vazquez,amaralemail}. Recent research has shown that such high variability in human behavior alters substantially the temporal dynamics of information diffusion and does not merely introduce some stochasticity in population-averaged models \cite{iribarren2009,karsai,miritello}. Thus, it is important to incorporate this human behavior into the models.

\smallskip

Besides,  information diffusion travels through social connections thereby depending on the properties of the social networks where it spreads. For example, simulations on synthetic scale-free networks showed that if information flowed through every social connection the epidemic threshold would be significantly lowered to the extent that it could disappear \cite{epidemic,satorras}, so that any rumor, virus or innovation might reach a large fraction of individuals in the population no matter how small the probability of being infected. Given the fact that social networks are scale-free \cite{siam} those results predict that there is a strong interplay between network structure and the spreading process. However such is not the case for information spreading processes. Our daily experience indicates that most rumors, innovations or marketing messages do not reach a significant part of the population \cite{wattsviral}. As mentioned earlier, the information transmission perceived cost prevents it from traveling inexpensively through all possible network paths. Therefore when participants assess the value of the information being passed, the impact of their social network structure on the diffusion process might be diminished. Unfortunately the true extent of such influence remains unknown in general. Moreover, the reach of information can be affected by the dynamics of human communication \cite{miritello} and thus it is important to understand the interplay between the static and dynamical properties of information diffusion.

\smallskip

Finally, there is an important shortcoming in the data currently available to investigate those questions. The vast majority of the large amount of data collected on information exchanges, for example email, SMS, calls or tweets, lacks the details required to follow the dynamics of a specific content item at the individual's level (see however \cite{debruyn}). Thus, the behavioral stochasticity of the individuals caused by the message content is masked and observations are limited to people's stochasticity due to the transmission media. A representative example of this difficulty is the study of communication patterns in mobile phone calls \cite{phonecalls,karsai,miritello} in which every communication, regardless of the message, is used to partially discover the social relationships network through which potential messages will spread but is not capable of revealing the specific dynamics of a particular piece of information. In other cases, data is not available at the individual participant level but just as population averaged metrics \cite{huberman,wattsviral} thereby hiding that different content items elicit diverse task prioritization in a given person or social segment. The situation is clearly unsatisfactory since, to our knowledge and possibly because of privacy concerns or data proprietorship, there are not very many data sets tracing the propagation of a specific piece of content throughout the social network (see however \cite{dwang}).

\smallskip

To overcome those limitations in the understanding of electronic information diffusion, we present here the results of a series of controlled Viral Marketing campaigns, the commercial form of WOM \cite{jurvetson}, that we conducted in eleven European countries. In them subscribers of a business online newsletter received incentives for recommending the newsletter subscription to their acquaintances. The detailed tracking of those recommendations revealed the factors impacting the diffusion dynamics of that particular piece of information at every step and suggested a branching process as the mechanism driving the dynamics of information diffusion. Thus the Bellman-Harris Branching Model, a generalization of the static percolation model introduced by Newman \cite{epidemic} for contagion propagation in networks, accurately describes our Viral Marketing campaigns. In particular, this branching model explains information diffusion of information in random networks and constitutes the simplest approach incorporating the human behavior high variability patterns both in activity volume and in response time.

\smallskip

The rest of this paper is organized as follows: Section \ref{sec:experiments} introduces our Viral Marketing campaigns and the information viral diffusion mechanism used in them, while Subsections \ref{sec:data} and \ref{sec:results}, respectively, present the campaigns propagation results data set and analyze the observed diffusion dynamics patterns and social connectivity found in such propagation. Section \ref{sec:model} follows with the analytical formulation of the Bellman-Harris Branching Model which includes detailed discussion of its phase transitions, asymptotic properties and time dynamics while Section \ref{sec:examples} studies several examples of its application to several scenarios of the response time distribution in the information propagation. We present our conclusions in Section \ref{sec:conclusions}. Finally, Appendix A discusses aspects of the substrate social network structure that can be gleaned through the information propagation process.

\section{Viral campaigns description}\label{sec:experiments}

We tracked and measured the ``Word-of-Mouth'' diffusion of viral marketing campaigns ran in eleven European markets that invited subscribers of an IT company online newsletter to promote new subscriptions among friends and colleagues. Campaign participants received incentives for spreading the offering through recommendation emails. The campaigns were fully web based. Banner ads, emails, search engines and the company web page drove participants to the campaign offering site. There, participants could fill in a referral form with names and email addresses of those to whom they recommended subscribing the newsletter. The submission of this form launched recommendation emails including a link to the campaign main page whose automatically generated URL was appended with codes allowing the web server to uniquely assign clicks on it to the sender and receiver of the corresponding email\footnote{Clicks on referral emails forwarded to a third person could not trace that individual and were assigned to their original receiver.}. The form, allowing up to four referrals per submission, checked destination email addresses for syntax correctness and to avoid self-recommendations. Cookies prevented multiple recommendations to the same address and improved usability by automatically filling-in sender's data in subsequent visits to the submission form. Additionally, the campaign server logged the time stamp of each step of the process (subscription, recommendation submission) and removed from records undeliverable recommendations.

\begin{figure}
\begin{center}
\includegraphics[width=0.45\textwidth,clip=]{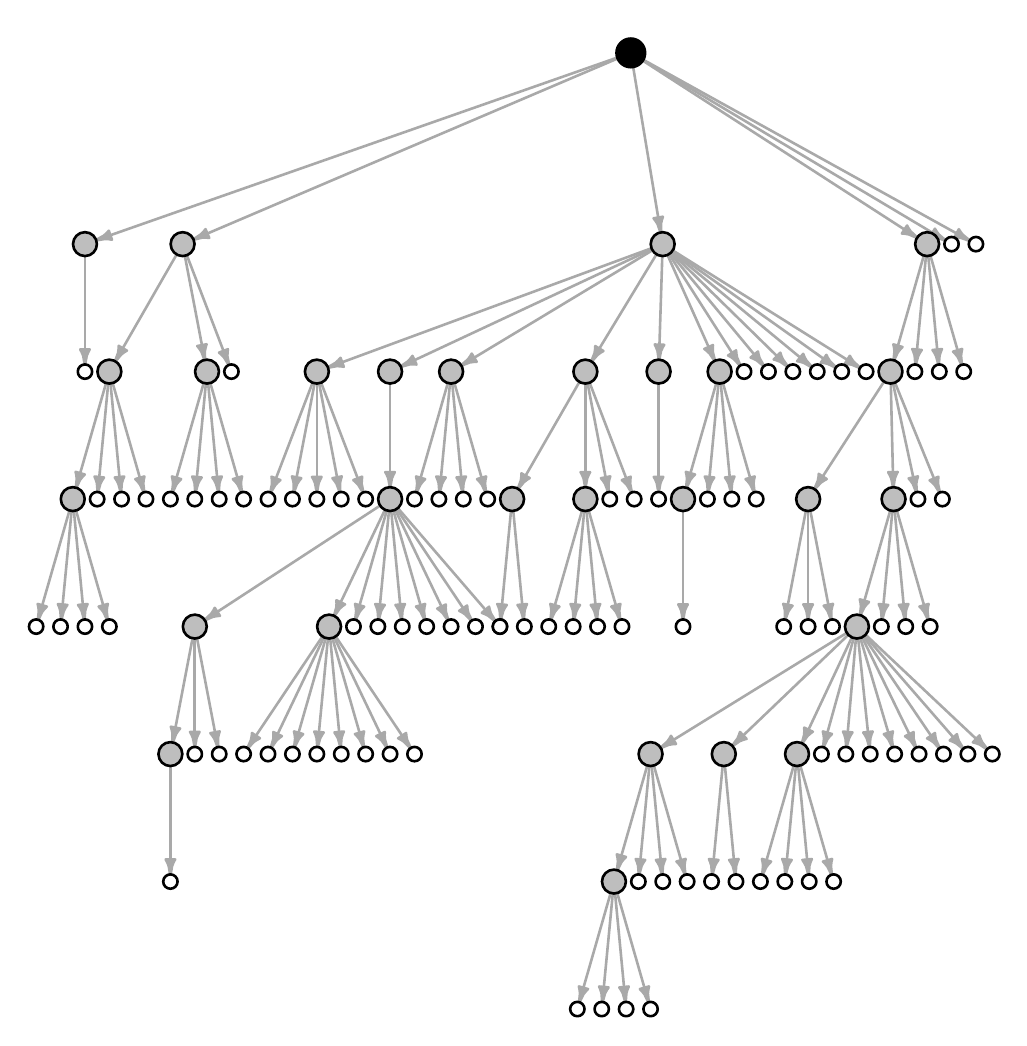}
\end{center}
\caption{The viral messages
diffusion graph of our campaigns is a set of 7,118 disconnected cascades like this one observed in Spain. Its 122 nodes (represented by dots) are grouped in 8 generations (horizontal layers) that stem from the generation zero node at the top (\emph{Seed} node, black) and grow through a branching process driven by the active nodes (gray) in each generation. Its tree-like structure is devoid of closed paths or triangles for a clustering coefficient $C=0$.}\label{cascade}
\end{figure}

\smallskip
\begin{table*}
\begin{tabular}{l|ccccc|ccccc|}
\hline
\textbf{Market} &	$N$	& $n_s$	& $\overline{s}$ & $s_{max}$ & $C_{cas}$ & $\overline{r}_{0}$ &	 $\overline{r}_{1}$ & $(\overline{r}_{1})_{SEM}$ & $\lambda_1$ &	 $R_1$ \\ \hline
France & 11,758	& 3,248	& 3.62 & 139 &	0.0000 & \,2.21 & 2.50	& 0.1023 & 0.062 & 0.154 \\	
DE+AT &	\,\,7,943   &	1,750       &	4.54    & 146 &	0.0049  & \,2.48     & 3.06  &	0.1155  &   0.092	 &	0.281 \\	 
Spain &	\,\,5,260	&\,\,\,\,843	&	6.24	&	122	& 0.0054 &	 \,3.16	 & 3.45  &  0.1909	&	 0.115	 &	0.397 \\	 
Nordic & \,\,2,509	&\,\,\,\,524	&	4.79	&\,\,\,34 &	0.0077 & \,2.82	 & 2.91  &	0.1836	&	 0.089	 &	0.259 \\	 
UK+NL &	\,\,2,111	&\,\,\,\,518	&	4.08	&\,\,\,25 &	0.0112 & \,2.49	 & 2.87  &	0.2398	&	 0.067	 &	0.192 \\	 
Italy &	\,\,1,602	&\,\,\,\,319	&	5.02	&\,\,\,41 &	0.0234 & \,2.87	 & 2.80  &  0.2301	&	 0.084	 &	0.236 \\	 \hline
\textbf{All markets } & \textbf{ 31,183} & \textbf{ 7,188}	& \textbf{ 4.34} & \textbf{ 146} & \textbf{0.0048} & \textbf{\,2.51} & \textbf{2.96} & \textbf{0.065} & \textbf{0.083} & \textbf{0.246}  \\ \hline
\end{tabular}
\caption{Structural and dynamic parameters of the viral diffusion network by market. Number of nodes ($N$) and of viral cascades ($n_s$), average cascade size ($\overline{s}=N/n_s$), largest cascade size ($s_{max}$) and Clustering coefficient of the \emph{Cascades Network} ($C_{cas}$). The diffusion dynamic parameters are the average number of recommendations sent by \emph{Seed} nodes ($\overline{r}_{0}$) or by \emph{Viral} nodes $\overline{r}_{1}$ (a.k.a. \emph{Fanout coefficient}) and the \emph{Transmissibility} $\lambda_1$. Also shown the \emph{Fanout coefficient} Standard Error of the Mean $(\overline{r}_{1})_{SEM}$ and Basic Reproductive Number $R_1$. Nordic comprises DK, FI, NO and SE.}\label{table2}
\end{table*}

The incentive to potential participants was the possibility of winning a laptop computer on a lottery taking place at the end of the campaign. The goal of such incentive was threefold: Firstly increasing participation, secondly, discouraging indiscriminate referrals which could lead to spamming-like behavior and, lastly, ensuring legal backup for tracking sender-receiver pairs as required by the campaign sponsor privacy policy. To reach those goals, eligibility to participate the lottery was limited to the so-called ``successful emails'' defined as any recommendation email whose recipient clicked on the coded URL included on it. Thus, the more referral emails sent to recipients who opened them and clicked their link, the bigger the sender's winning odds. The lottery draw was held among successful recommendations only and both sender and receiver of the winning recommendation would receive the prize. The campaign terms and conditions, accessible from all web pages, stated that participation in the prize draw implied the sender's and receiver's authorization for the system recording the details of their email transaction since it was necessary to ensure that both parties could receive the prize if their email was a winner. Subscribing to the newsletter was not required to take part in the prize draw. Campaigns in all countries ran in local language but were identical otherwise: Same offering, incentive, eligibility rules, prize draw mechanism, campaign period, web user interface and tracking processes. This ensured equivalence of the experiment in all countries and allowed tracing differences in observed behavior to the market specifics and not to the campaigns execution. In addition, this guaranteed the neutrality of the messages content in regards to the recipients' reaction. Unsuccessful emails, disconnected nodes, nodes with invalid or undeliverable email addresses, self-recommendations and multiple referrals between same nodes were discarded. The message viral propagation network was built from such cleansed data set and its key parameters measured with standard network analysis tools. Personal information was encrypted to protect the participants' privacy.

\subsection{Campaigns propagation data set}\label{sec:data}

Spurred by the sponsor web sites, email marketing and exogenous online advertising, a total of 7,225 individuals acted as \emph{Seed} nodes by initiating message diffusion cascades which subsequently grew through viral pass-along driven by 2,002 \emph{secondary spreaders} which we will also designate as \emph{Viral} nodes in what follows. Thus the viral offering touched another 21,956 individuals who did not forward it and were, therefore, passive nodes. All in all, and as shown in Table \ref{table2}, a total $N=31,183$ individuals, of which 9,227 were active spreaders, received the viral message. Thus, $77\%$ of the campaigns participants received the message through the endogenous viral propagation mechanism. The 7,188 tree-like, independent propagation cascades originated by this process such as the one in Fig. \ref{cascade}, form the \emph{Cascades Network}, a sparse graph whose nodes representing campaign participants are connected by 24,207 directed links formed by the recommendation emails they sent. Besides, the viral cascades are generally almost pure trees, with very few loops or closed triangles, as evidenced by the Clustering Coefficient of the network of all markets $C_{cas}=0.0048$, which is two orders of magnitude lower than typical values reported for social networks \cite{why}: for example $C_{eml} = 0.156$ measured in a typical email network of similar size \cite{ebelemail}.

\smallskip

By analogy to the spreading of diseases \cite{andersonmay}, diffusion of information in a population is often described by average quantities. Although receiving and propagating messages can be quite involving processes, population-level analysis describes information propagation as a function of the probability $\lambda_1$ of a person becoming \emph{secondary spreader} after receiving a message from a \emph{Seed} node and of the average number of people $\overline{r}_1$ contacted by such {\em secondary spreaders}. In this simple approach those two parameters, \emph{Transmissibility} ($\lambda_1$) and \emph{Fanout coefficient} ($\overline{r}_1$), fully characterize the mean-field description. In our campaigns only $8.36\%$ of the participants receiving a recommendation email from other participant engaged in spreading it themselves and thus $\lambda_1 = 0.0836$. Those secondary spreaders sent, in average, $2.96$ messages each and hence the \emph{Fanout coefficient} was $\overline{r}_1=2.96$. Interestingly, this value is higher than the average number of recommendations ($\overline{r}_0 = 2.51$) sent by the \emph{Seed} nodes that triggered cascades after becoming aware of the campaign message through market seeding tactics. Such gap stems from the combination of two factors: Firstly, a stronger involvement in the diffusion of the individuals receiving the message from a trusted source versus those who found the campaign by chance \cite{trust} and, secondly, the ``Friendship paradox'' \cite{feld}, a property of networks which causes individuals reached trough messages sent by others to be more connected in average than those chosen at random: for example, in a random network with node degree distribution $P(k)$ the probability of randomly picking a node of degree $k'$ is $P(k')$ whereas the probability of a message coming from any node reaching a node of degree $k'$ is $k'P(k')/\overline{k}$, bigger than $P(k')$ for $k'>\overline{k}$ \cite{siam}. Thus highly connected nodes are more likely to be reached by messages already spreading through the network than by exogenous marketing tactics (web banners or email tactics) which do not benefit from such network effect when the message spreading starts. This phenomenon causes \emph{secondary spreaders} to have more contacts in average than \emph{Seed} nodes and more choices to forward the message.

\smallskip
\begin{table}
\begin{tabular}{l|rcccccc|}
\hline
Node class &\ \emph{N} &\ $\overline{r}$ &\ $\overline{r^2}$ &\ $\sigma_r$ &\ $(\overline{r})_{SEM}$ &\ $\alpha$ &\ $\beta$ \\ \hline
Seed (0) &\  7,225 &\  2.51 &\ 15.14 &\ 2.97 &\ 0.035 &\ 3.50 &\ \,\,30.52 \\
Viral (1) &\  2,002 &\  2.96 &\ 18.10 &\ 3.05 &\ 0.068 &\ 3.71 &\ 100.88 \\ \hline
Active (a) &\  9,227 &\ 2.61 &\ 15.82 &\ 3.00 &\ 0.031 &\ 3.54 &\ \,\,39.48 \\ \hline
\end{tabular}
\caption{Statistics of the viral campaigns participants recommendation activity $(r)$ by node class. Active nodes $(a)$ are the union of the \emph{Seed} $(0)$ and \emph{Viral} $(1)$ classes. The probability distribution of the number of recommendations $(r)$ fits a Harris power-law of the form of Eq. (\ref{harris}) with $\alpha$ and $\beta$ estimated by the method of moments using $\overline{r}$ and $\overline{r^2}$.}\label{table1}
\end{table}

On the other hand, it makes sense to assume that the number of recommendations sent by \emph{secondary spreaders} (including not sending any) results from a decision by each message recipient that involves a trade-off between the message forwarding cost and its perceived value. For our campaigns lottery prize for example, and in a population average approach, a reasonable proxy of the perceived value of winning the prize for residents in a given country could be the fraction of the average income of its citizens represented by the prize cost in that market. Granted, there may be many other factors at play in the formation of such perception, but there is a very significant correlation ($\rho=0.6$) between the average income and the average number of recommendations $\overline{r}_1$ sent by \emph{secondary spreaders} in each market which indicates that the expected gain average relative size may be one of them (see Table \ref{table2}).

\smallskip

Additionally, the human intervention in such decision process is at the root of a very unique property of the dynamics of information diffusion. Comparing viral campaigns parameters in different markets (see Table \ref{table2}), we observe a wide range of values in their respective information propagation dynamical parameters. Since the campaigns execution was identical in all markets, those variations can only be due to a change in perception of the viral offering value and/or of the message forwarding cost by customers in each market. Interestingly, variations of the \emph{Transmissibility} ($\lambda_1$) and the \emph{Fanout coefficient} ($\overline{r}_1$) present a Pearson coefficient $\rho=0.92$ as evidence of a very strong dependence between them. We proved in \cite{affinity} that such dependence has the form
\begin{equation}\label{correlation}
\overline{r}_1=1+b(1-e^{-c\lambda_1}), \qquad 0\leq\lambda_1<1
\end{equation}
which reduces to $\overline{r}_1\simeq1+a\lambda_1$ $(a=bc)$ for $c\lambda_1< 1$. This peculiarity of information diffusion processes, not observed in disease epidemics, arises because the decisions of becoming a spreader and of the number of viral messages to send are simultaneously made by each participant which introduces correlation in their averages.

\subsection{Diffusion dynamics analysis}\label{sec:results}

\begin{figure}
\centering
\includegraphics[width=8.2cm,clip=]{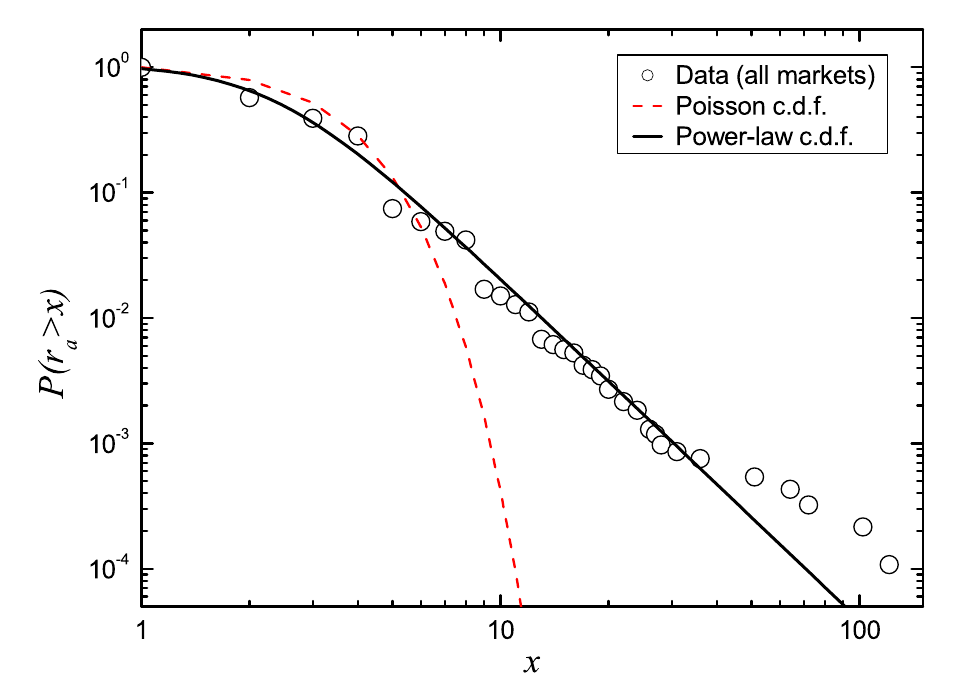}
\caption{(Color online) Active nodes (\emph{Seed}+\emph{Viral}) cumulative probability distribution for campaigns in all markets (circles). Solid line is the fit to a power-law $P(r) = H_{\alpha\beta}/(\beta+r^\alpha)$ whose pdf exponent is $\alpha = 3.54\pm0.02$ (see Table \ref{table1}). Dashed line is the prediction of a discrete Poisson distribution with the mean of the empirical data ($\overline{r}_a = 2.61$).}
\label{fig1b}
\end{figure}

In a first approximation we could analyze information dynamics by studying the Basic Reproductive Number $R_1$ of epidemiology, the average number of secondary cases generated by each virally informed individual, which results from the definition of the dynamical parameters as $R_1 = \lambda_1 \overline{r}_1$. However, average quantities like $R_{1}$ hide the heterogeneous nature of epidemics \cite{ssdisease} and also of information diffusion. In fact our campaigns show that most of the observed transmission occurs due to extraordinary events. In particular, we get that the probability distribution function (pdf) of the number of recommendations sent is well approximated by the Harris discrete distribution
\begin{equation}\label{harris}
p_{r}=\frac{H_{\alpha\beta}}{\beta+r^{\alpha}}, \qquad r=1,2,\ldots
\end{equation}
where $H_{\alpha,\beta}$ is a normalization constant so that $\sum_{r=1}^{\infty}p_{r}=1$. This function displays a power-law behavior $p_{r} \sim r^{-\alpha}$ in its tail starting approximately at the cutoff point $r^*\simeq \beta^{1/\alpha}$. Table \ref{table1} lists the distribution  parameters for \emph{Seeds} ($p_{0,r}$), \emph{Viral} ($p_{1,r}$) and total \emph{Active} ($p_{a,r}$) nodes while Fig. \ref{fig1b} shows the probability distribution  of the recommendations sent by \emph{Active} nodes in all markets, and the comparison to the probability predicted by a Poisson discrete distribution with mean $\overline{r}=2.61$, same as that of the empirical data. The markedly different behavior between both of them indicates the high probability of finding individuals making a large number of recommendations. As noted in the introduction, such high demographic stochasticity, observed in many other human activities \cite{barabasinature,telephone,blogs,blogs1,pitkow,tipping,sexual}, suggests that humans' response to a particular task cannot be described by close-to-average models where they are all assumed to behave in a similar fashion with some small degree of demographic stochasticity \cite{bass}. In sharp contrast with population homogeneous models of information spreading, we found that 2\% of the active population in our viral campaigns has $r_a>10$ suggesting the existence of super-spreading individuals.

\smallskip

\begin{figure}
\centering
\includegraphics[width=8.2cm,clip=]{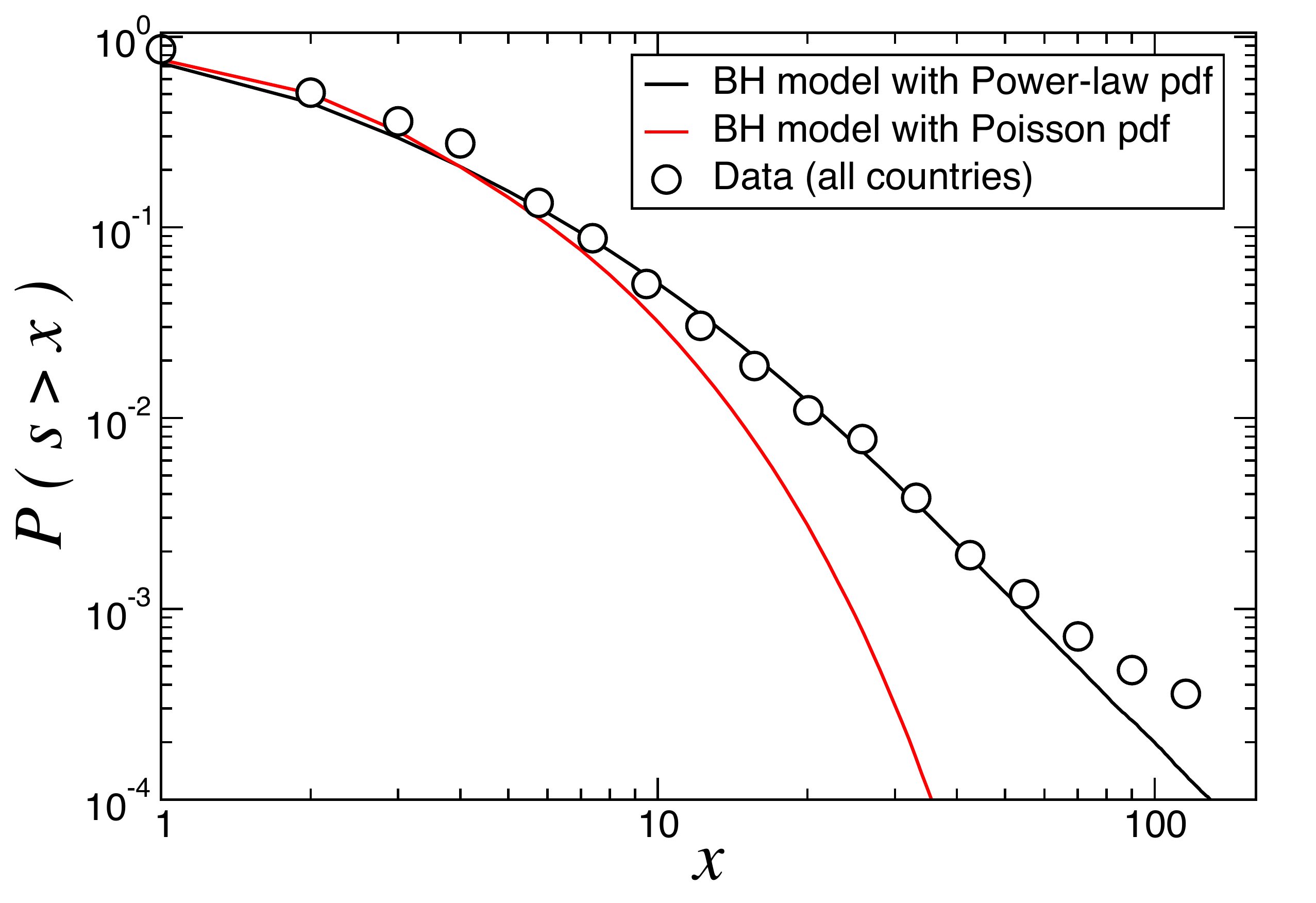}
\caption{(Color online) Cumulative distribution of viral cascades size in all markets (circles). The power-law line underneath the circles is not a fit of the data but the prediction of the Bellman-Harris branching model with a power-law pdf for the recommendations distribution. The line below it is the branching model prediction with a Poisson distribution (see Sec. \ref{sec:model}).}\label{casc}
\end{figure}

Super-spreading individuals have also been found in non-sexual disease spreading \cite{ssdisease} where they significantly increase outbreak sizes. In a similar manner, the sizes of the information cascades found in our campaigns indicate that super-spreading individuals are responsible for making large viral cascades rarer but more explosive. The probability distribution of the campaigns cascades sizes, represented (see Fig.~\ref{casc}), is also a fat-tailed distribution (in fact, the tail can be fitted to a power law $p_s \sim s^{-\beta}$ with $\beta \simeq 3.2$). In contrast, neglecting the existence of super-spreading individuals but still considering some degree of stochasticity in the number of recommendations by assuming $p_{a,r}$ is a Poisson distribution with the same average, a cascade like the one in Fig. \ref{cascade} would have an occurrence probability of approximately once every $10^{12}$ {\em Seed} nodes, a number much larger than the total world population (see Fig. \ref{casc}).

\smallskip

An element to consider in the aforementioned spreading stochasticity is the impact, if any, of the underlying social network heterogeneity in a similar way to that of the connectivity of a computer network on the diffusion of computer viruses \cite{NM}. Social networks data reveals that humans show large variability in their number of social contacts \cite{scalefree}. Thus, the connectivity $k_i$ of email networks whether measured by email traffic or by the users' email address books is fat-tailed distributed \cite{NM,ebelemail}. In some cases it is power-law distributed like the number of recommendations in our campaigns. Large variability in the numbers of social contacts has a deep effect on disease spreading \cite{epidemic,satorras}. In fact, disease spreading models on networks show that if information flows with the same probability through any link in a social network, its topological properties can significantly lower the ``tipping-point''\footnote{Epidemiology term designating the point in a contagion process where its spreading rate increases dramatically and changes the nature of the process.}. However, while indiscriminate propagation can happen in computer viruses, diseases or other mechanistic processes, the human handling of information diffusion limits the influence of the social network structure: we expect, in general, the number of recommendations to be small compared to the social connectivity ($r_i \ll k_i$). While in social networks the ``Friendship paradox'' \cite{feld,siam} implies that $\overline{k}_{nn} \gg \overline{k}$ (with $\overline{k}_{nn}$ the average number of social contacts of an individual's neighbors and $\overline{k}$ the average number of social contacts of an individual), our recommendation network features $\overline{r}_{nn} \equiv \overline{r}_v \simeq\overline{r}_s$. If, as supposed in most models \cite{yamir,satorras}, information flows through a fraction of the social contacts of an individual, we should have $\overline{r}_{nn} \gg \overline{r}$ instead. A way to recover our result is to assume that $r_i$ and $k_i$ are largely independent. Our tree-like diffusion cascades lead to a low undirected clustering coefficient \cite{siam} of the viral \emph{cascades network} ($C_{cas} = 0.048$) compared to the values reported for email social networks ($C_{soc} \sim 0.15 - 0.25$) \cite{NM} which supports such assumption. Assuming $r_i$ and $k_i$ independent, we get (App. \ref{appA})
\begin{equation}\label{eq:cvir}
C_{cas} \sim \frac{2 R_{1}}{(\langle\overline{k}_{nn}\rangle-1)}\times C_{soc}, \qquad R_1 \ll 1
\end{equation}where $\overline{k}_{nn}$ is the average number of social contacts of the neighbors of an individual. In social networks $\langle \overline{k}_{nn} \rangle$ is a large number which leads to a very low clustering coefficient even for processes close to the ``tipping-point'' ($R_{1} \simeq 1$). This fact explains the unreasonable effectiveness of tree-based theory to explain information diffusion on networks with clustering \cite{melnik}. In conclusion, large heterogeneity of recommendations activity is due to the participants' behavior rather than consequence of their connectivity degree which is just the activity upper bound.

\smallskip

\begin{figure}
\centering
\includegraphics[width=0.45\textwidth,clip=]{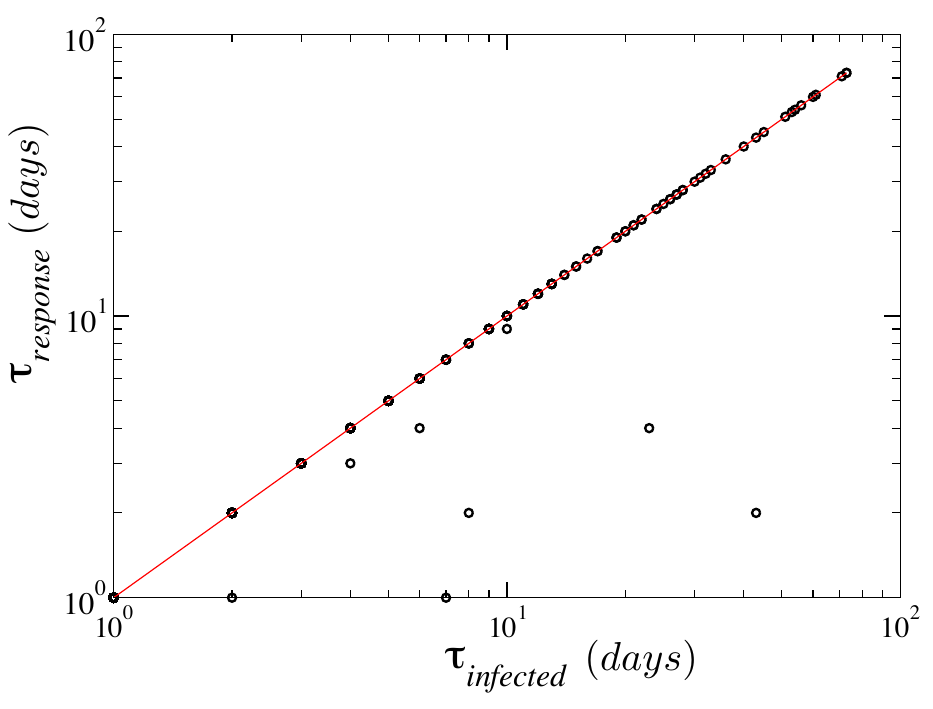}
\caption{(Color online) Relationship between the time (days) a \emph{Viral} node remains active (infected) $\tau_{infected}$ and the time elapsed until it resends its first message ($\tau_{response}$) for each \emph{Viral} node in our campaigns. The line is $\tau_{infected}=\tau_{response}$. Nodes on the line sent the message to all contacts at once while those outside it remained spreaders for a longer period. Only early responders ($\tau_{response}<10$ days) have some likelihood of staying active for more than one forwarding session.} \label{figti}
\end{figure}

Finally, another important aspect to consider in the dynamics of information diffusion is the nodes' reaction to receiving a message: Shall they decide to spread it, how long do they take to do so?, for how long do they remain active?, and, is their responsiveness correlated in any way to the number of contacts they resend the message to? The answer to these questions lies in the increasing evidence that the timing of many human activities, ranging from communication to entertainment and work patterns, follow non-Poisson statistics, characterized by bursts of rapidly occurring events separated by long periods of inactivity \cite{barabasinature}. In fact, our campaigns revealed that most of the active nodes turn inactive right after spreading the information once which means that \emph{Viral} nodes do not remain as spreaders for a long time. The top panel in Fig. \ref{figti} shows that for most of the \emph{Viral} nodes (actually 97\% of them), the lapse of time between receiving the message and passing it along $\tau_{response}$ equals the interval between receiving the message and the last time it has been resent $\tau_{infected}$. The fact that for the most part \emph{Viral} nodes show just one spreading event means, from a modeling perspective, that diffusion follows an almost pure ``birth and death'' model. Besides, the time dynamics of the viral recommendation process is independent from the number of recommendations $r_1$ sent by \emph{Viral} nodes as was shown in \cite{iribarren2009}, that is there is no correlation between such number and the response time $\tau_{response}$ as evidenced by the Pearson correlation coefficient of the two variables ($\rho=-0.05$). As we have shown in \cite{iribarren2009}, the probability distribution function of the \emph{Viral} nodes response time $P(\tau_{response})$ is a long tailed log-normal in another evidence of the humans' large heterogeneity in WOM diffusion. In this sense, participants behave like a SIR model in which infection and decay to the recovered state happen at the same time \cite{andersonmay}.

\section{Branching Dynamics Model}\label{sec:model}

\begin{figure*}
\centering
\includegraphics[width=0.8\textwidth,clip=]{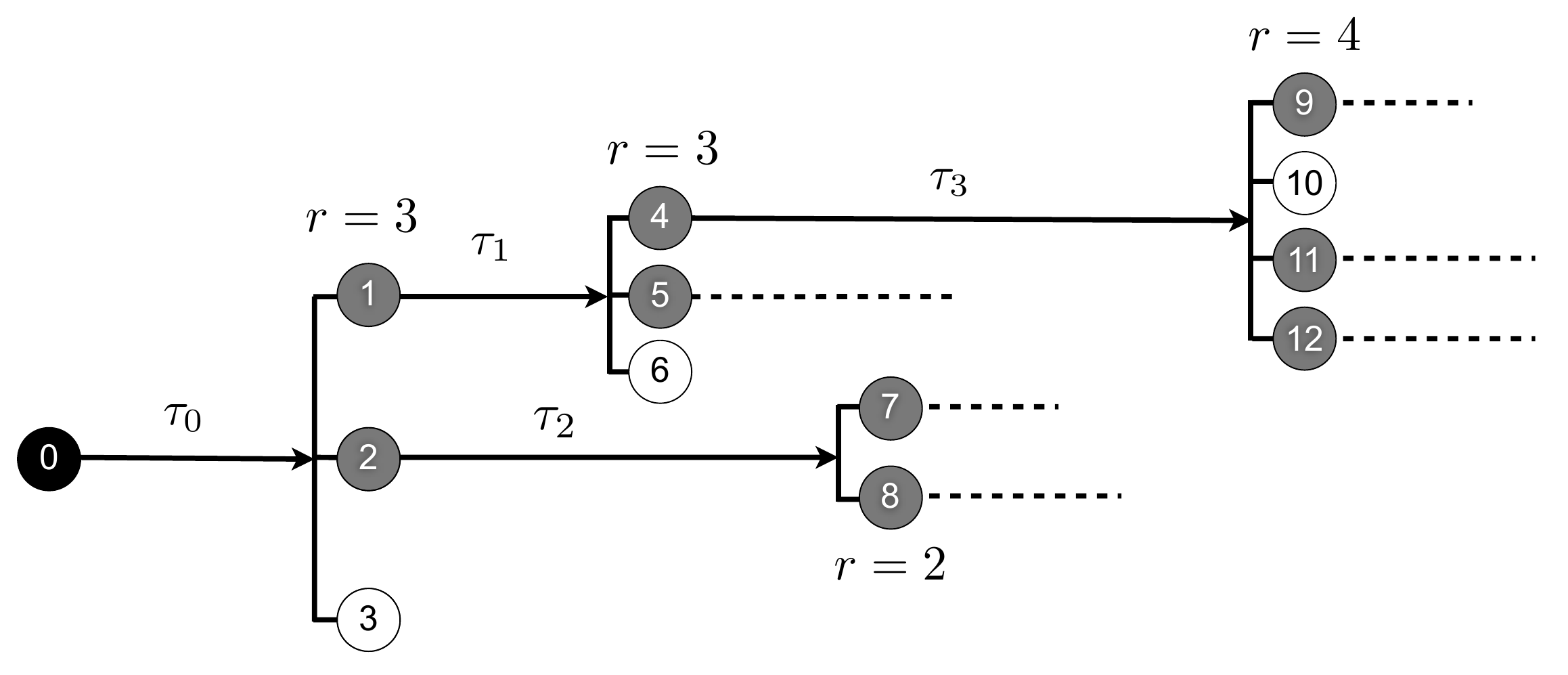}
\caption{Flowchart example of cascades generated by the Bellman-Harris branching model used to explain diffusion of information in social networks: the cascade starts with a \emph{Seed} (labeled $0$) which sends the information to $r = 3$ of its social contacts after time $\tau_0$. \emph{Viral} nodes 1 and 2 are ``infected'' and forward the message to $r = 3$ and $r = 2$ social contacts after times $\tau_1$ and $\tau_2$ respectively, while uninterested node 3 remains inactive. Values of $r$ and $\tau$ are independent and sampled from distributions $P(r)$ and $G(t)$. Propagation continues until there are no active nodes left. Time increases left to right.}\label{figbh}
\end{figure*}

\smallskip

The study of our experimental data leads to a theoretical framework for the process of information diffusion where the dynamics of information viral spreading is explained by tree-like cascades. Each information cascade stems from an initial \emph{Seed} that starts the viral message propagation with a random number of recommendations distributed as $p_{0,r}$ and whose average is $\overline{r}_{0}$. The individuals reached by the message become \emph{secondary spreaders} with probability $\lambda_1$ thereby giving birth to a new generation of \emph{Viral} nodes which, in turn, propagate the message further with $r_{1}$ recommendations distributed by $p_{1,r}$ with average $\overline{r}_{1}$. After sending their recommendations individuals become inactive and the process continues stochastically through new individuals in successive generations until none of the members of the latest one spread the message. At that point the information cascades die out and the propagation ends. This process corresponds to the well known Bellman-Harris (BH) branching model \cite{branching,athreya,iribarren2009} which is the simplest mathematical framework to study the branching dynamics of information diffusion. It generalizes the static and Markovian Galton-Watson model typically used to model information diffusion \cite{iribarren2009,dwang,golub,Lans2010} or, in general, percolation processes in social networks \cite{epidemic}.

\smallskip

In the BH model, those two distributions, $p_{0,r}$ and $p_{1,r}$ ($r_i=1,2\ldots$), represent the number of recommendations sent by \emph{Seed} and \emph{Viral} nodes respectively. The introduction of two different distributions for the recommendations sent by \emph{Seed} and \emph{Viral} nodes is not only due to the difference in the average number of recommendations observed in our campaigns (see table \ref{table1}) but also because, in general, in social networks the average connectivity of a node's nearest neighbors is higher than the average connectivity of the network nodes themselves. In particular, for completely uncorrelated random networks with distribution of connectivity given by $P(k)$ the distribution of the number of connections of the nearest neighbors of a node is  $P'(k) = k P(k)/\overline{k}$ \cite{dorogovtsev}. The case in which informed nodes decide not to pass along the information can be incorporated in the recommendations distribution as the case in which the number of messages sent is $r_1 = 0$. Thus we can construct a family of probability distributions of the recommendations sent by nodes $\tilde p_{i,r}$ where
\begin{equation}
\tilde p_{i,0} = (1-\lambda_i),\qquad \tilde p_{i,r} = \lambda_i p_{i,r}\quad r\geq0
\end{equation}
from whence one can obtain the average number of recommendations in the new distributions which are related to the primary and secondary reproductive numbers as
\begin{eqnarray}
\sum_{r\geq 0} \tilde p_{0,r} r= \lambda_0 \overline{r}_0 = R_0 \\
\sum_{r\geq 0} \tilde p_{1,r} r = \lambda_1 \overline{r}_1 = R_1
\end{eqnarray}

To formalize the study of the information spreading branching process, we define now the generating functions
\begin{equation}
f_0(x) = \sum_{r=0}^{\infty} \tilde p_{0,r} x^{r},\qquad f_1(x) = \sum_{r=0}^{\infty} \tilde p_{1,r} x^{r}
\end{equation}

Moments of the $\tilde p_{i,r}$ distributions can be obtained through derivatives of the generating functions
\begin{equation}
R_0 = f'_0(1),\quad \sigma_0^2 = f''_0(1) + f'_0(1) - [f'_0(1)]^2
\end{equation}
where $\sigma_0^2$ is the variance of the number of recommendations of \emph{Seed} nodes. We will also assume different cdf of response times ($\tau_{infected}$) for \emph{Seed} and \emph{Viral} nodes which we will denote as $G_{0}(t)$ and $G_1(t)$. Their means are $\overline{\tau}_0$ and  $\overline{\tau}_1$ respectively.

\smallskip

We want to determine the probability distribution of finding $I(t)$ nodes active (i.e. recommending) at time $t$ provided we start with one participant at $t=0$, i.e. $I(0) = 1$. To do that we use the following self-consistent argument: since the number of recommendations sent by each \emph{Viral} node are random independent processes, the branching process starting from each \emph{Viral} node after a given recommendation, which we denote $I_1(t)$ (with $I_1(0) = 1$) are independent identically distributed (iid) copies of the same process. For example, in Fig. \ref{figbh} the branching process starting from nodes 1 and 2 are iid copies of the same process $I_1(t)$. But also, the $I_1(t)$ process starting from 1 and the $I_1(t)$ processes starting from 4 and 5 must be statistically the same. Thus we have a self-consistent relationship between the branching process starting at a \emph{Viral} node and the processes starting from any of its $r_1$ recommendations:
\begin{equation}
I_{1}(t) = \left\{
\begin{array}{ll}
1 & if\ t < \tau\\
\sum_{i=1}^{r_1} I_{1}^{(i)} (t-\tau) & if\ t\geq\tau
\end{array}
\right.
\end{equation}
where $I_{1}^{(i)}(t)$ are iid copies of the branching process $I_{1}(t)$ and assuming that the recommendation event happens at $t = \tau$. Note that in this self-consistent equation $r_1$ (the number of recommendations made by a \emph{Viral} node and distributed by $\tilde p_{1,r}$) and the time $\tau$ are both random and independent. To describe the process we use generating functions techniques: we define the generating function for $I_1(t)$ as  $F_{1}(s,t) = \sum_{k\geq 0}P[I_{1}(t) =k] s^{k}$, and thus we get
\begin{equation}\label{eq1f1}
F_{1}(s,t) = \left\{
\begin{array}{ll}
s & if\ t < \tau\\
f_1[F_{1}(s,t-\tau)]  & if\ t\geq \tau
\end{array}
\right.
\end{equation}

Finally, since $\tau$ occurs randomly with cdf $G_{1}(\tau)$, one can integrate equation (\ref{eq1f1}) over $\tau$ to get
\begin{equation}\label{eqf1}
F_{1}(s,t) = s [1- G_{1}(t)] + \!\! \int_{0}^{t}\!\!  dG_{1}(\tau) f_1[F_{1}(s,t-\tau)]
\end{equation}

The same reasoning can be applied to the \emph{Seed} nodes, with the exception that now the number of recommendations are distributed according to $\tilde p_{0,r}$. Denoting $I_{0}(t)$ the process starting from an initial seed then we have
\begin{equation}\label{eq1f0}
I_{0}(t) = \left\{
\begin{array}{ll}
1 & if\ t < \tau\\
\sum_{j=1}^{r_0} I_{1}^{(j)} (t-\tau) & if\ t\geq \tau
\end{array}
\right.
\end{equation}
where once again $I_{1}^{(j)}(t)$ are $j$ copies of the branching process $I_{1}(t)$ and $r_0$ is a random number with probability distribution $\tilde p_{0,r}$. The same reasoning above leads to
\begin{equation}\label{eqf0}
F_{0}(s,t) = s [1- G_{0}(t)] +\! \! \int_{0}^{t}\!\!  dG_{0}(\tau) f_0[F_{1}(s,t-\tau)]
\end{equation}

This equation is the one that describes the time dynamics of our branching process, starting from a given \emph{Seed}. Note that it is a non-homogeneous equation, since it depends on the solution of Eq. \ref{eqf1}. Thus we must first try to solve Eq. \ref{eqf1} and then insert its solution in Eq. \ref{eqf0}.

\smallskip

Identical reasoning can be used to derive the equations for $S_0(t)$ [$S_1(t)$], the size of a cascade at time $t$ starting from a \emph{Seed} or \emph{Viral} node at $t=0$ to obtain
\begin{equation}\label{eq1f0}
S_{0}(t) = \left\{
\begin{array}{ll}
1 & if\ t < \tau\\
1+\sum_{i=1}^{r_0} S_{1}^{(i)} (t-\tau) & if\ t\geq \tau
\end{array}
\right.
\end{equation}
where
\begin{equation}
S_{1}(t) = \left\{
\begin{array}{ll}
1 & if\ t < \tau\\
1+\sum_{i=1}^{r_1} S_{1}^{(i)} (t-\tau) & if\ t\geq \tau
\end{array}
\right.
\end{equation}

Thus, the generating function for the cascade sizes
\begin{eqnarray}
\Phi_0(s,t) &=& \sum_{k=1}^\infty P[S_0(t) = k] s^k \\
\Phi_1(s,t) &=& \sum_{k=1}^\infty P[S_1(t) = k] s^k
\end{eqnarray} are the solution of the integro-differential equations
\begin{equation}\label{eqs0}
\Phi_0(s,t) = s[1-G_0(t)] + s\!\! \int_0^t \! \! dG_0(\tau)f_0[\Phi_1(s,t-\tau)]
\end{equation}
\begin{equation}\label{eqs1}
\Phi_1(s,t) = s[1-G_1(t)] + s\!\!  \int_0^t \! \! dG_1(\tau)f_1[\Phi_1(s,t-\tau)]
\end{equation}

Note that these equations generalize the static ones introduced by Newman \cite{epidemic} and include the example of epidemics in configuration model networks in \cite{Karrer2010}. General solutions for equations (\ref{eqf1}), (\ref{eqf0}), (\ref{eqs0}) and (\ref{eqs1}) are not known, but some special cases and limits can be studied. In the following subsections, we study some properties of the model and compare its predictions with our experiments and other theoretical situations.

\subsection{The ``tipping-point''}

We are interested in the dynamical process when time is large enough, but also in the asymptotic regime when $t \to \infty$. In particular, the overall probability $q$ of extinction of the cascade is given by the probability that the initial \emph{Seed} does not propagate the information $(1-\lambda_0)$ and that, even when the \emph{Seed} propagates the infection to some nodes, the branches stemming of the eventual \emph{Viral} nodes die out. In this case the extinction  probability $q_1$ of a branch starting by a \emph{Viral} node, i.e. the probability of $I_1(t)=0$ (number of new nodes in the branch) for any finite time $t$, results from the generating function as
\begin{equation}
q_1 = \lim_{t \to \infty} P[I_1(t) = 0] = \lim_{t \to \infty} F_1(0,t) \equiv F_1(0,\infty)
\end{equation}

Inserting this definition in equation (\ref{eqf1}) we get that $q_1$ is the root of
\begin{equation}\label{extprob}
q_1 = f_1(q_1).
\end{equation}

Since generating functions are convex and $f_1(1) = 1$ we get that if $R_1 = f_1'(1) \leq 1$ the only solution is $q_1=1$, while if $R_1 > 1$ there exists a solution $0 \leq q_1 < 1$. The point $R_1 = 1$ is known as the ``tipping-point'', since above it there is a finite probability $1-q_1$ that a viral cascade does not die out and thus grows infinitely, while below the ``tipping-point'' $q_1=1$ and thus every cascade started by a \emph{Seed} node will eventually die out. Including the probability of \emph{Seeds} not making any recommendation we obtain the probability that a cascade dies out
\begin{equation}
q_0 = 1-\lambda_0 + \lambda_0 q_1
\end{equation}
and using the results for $q_1$ we get that $q_0 = 1$ below the ``tipping-point'' and $q_0 < 1$ above the ``tipping-point''. For our campaigns \emph{Seeds} are active by definition which means that $\lambda_0=1$ and $q_0=q_1=1$ in all cases.

\begin{figure}
\centering
\includegraphics[width=0.41\textwidth,viewport= 120 120 2000 1580,clip=]{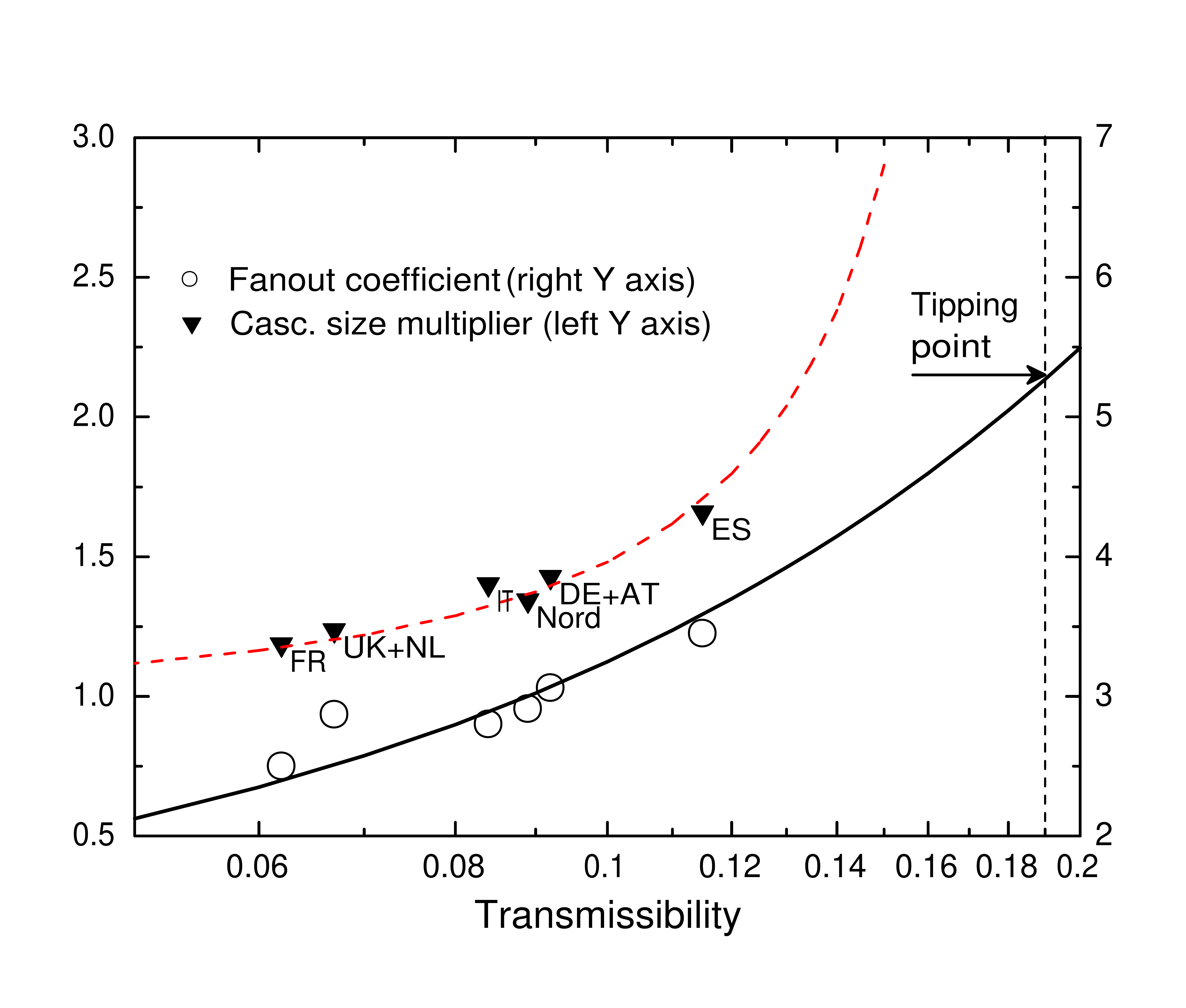}
\caption{(Color online) Size of viral cascades as a function of $\lambda_1$ for markets in Table \ref{table2}. Triangles represent the cascade size multiplier $1/(1-R_1)$ (left Y axis). Dashed line is not a fit but the prediction of the branching model (Eq. \ref{clustersize}) which diverges at the ``tipping-point'' ($\overline{r}_1 \simeq 5.27$, $(\lambda_1)_c \simeq 0.19$) estimated by the correlation $\overline{r}_1=1+22.48\lambda_1$ existing between \emph{Transmissibility} and \emph{Fanout coefficient} (solid line, circles, right Y axis).} \label{tipping}
\end{figure}

\smallskip

Moreover, using the correlation between $\lambda_1$ and $\overline{r}_{1}$ in Eq. (\ref{correlation}) and the condition $R_1=(\lambda_1)_c\overline{r}_1=1$ one can estimate the critical viral transmissibility $(\lambda_1)_{c}$ required for the viral message to percolate through a large fraction of the entire network. We obtained that $(\lambda_1)_{c}=0.19$ which corresponds to $(\overline{r}_{1})_c = 5.27$. Of course this is an upper limit to the real ``tipping-point'' since it is based on the assumption that cascades originating from different {\em Seeds} do not merge as the propagation progresses which is only valid far from the ``tipping-point''. The low average number of recommendations needed to attain the ``tipping-point'' illustrates the limited effect of the social network topology on the viral campaigns efficiency: It is not necessary to forward the message to each participants' social contact in order to reach a significant fraction of the network population. Fig. \ref{tipping} shows the estimation of our campaigns message propagation ``tipping-point'' based on such findings. While both $\lambda_1$ and $\overline{r}_1$ vary with the market where the campaign ran (see Table \ref{table2}) we found that $R_{1}<1$ for all cases, i.e. the viral propagation did not reach the ``tipping-point''.

\subsection{Asymptotic properties}\label{asymp}

As we have seen, below the ``tipping-point'' $q_1=1$, that is, all viral cascades die out eventually. This means that there must exist an asymptotic distribution for the size of the cascades $\Phi(s,\infty) = \lim_{t\to\infty} \Phi(s,t)$ which is the solution of equations (\ref{eqs0}) and (\ref{eqs1}) in the limit $t\to \infty$
\begin{eqnarray}
\Phi_0(s,\infty) & = & s f_0[\Phi_1(s,\infty)]\\
\Phi_1(s,\infty) & = & s f_1[\Phi_1(s,\infty)].
\end{eqnarray}

These equations were obtained previously by Newman \cite{epidemic}. In particular we can obtain the average and variance of the cascades size by using $\langle S_0(\infty) \rangle = \Phi_0'(1,\infty)$ and $\mathrm{Var}[S_0(\infty)] =\Phi_0''(1,\infty) + \Phi_0'(1,\infty) - [\Phi_0'(1,\infty)]^2$ to get
\begin{eqnarray}
\langle S_0(\infty) \rangle&=&1 + \frac{R_0}{1-R_1} \label{means} \\
\mathrm{Var}[S_0(\infty)] &=& \sigma_0^2R_1^2 + R_0\frac{\sigma_1^2+R_1^2}{1-R_1^2} \label{vars}
\end{eqnarray}

As expected, when we approach the ``tipping-point'', $R_1 \to 1$, the average and variance of the cascade size diverges. With $\lambda_0=1$ in eq. (\ref{means}) we get the following expression for the average cascade size at infinite time
\begin{equation}\label{clustersize}
\overline{s^*} = 1 + \frac{\overline{r}_{0}}{1-R_1}, \qquad 0\leq R_1<1
\end{equation}
which, using the parameters for all markets in Table \ref{table2}, estimates the average cascade size in our campaigns as $\overline{s*}=4.4$, very close to the observed value ($\overline{s}=4.34$). Not only are average cascade sizes well predicted by the branching model, but their distribution, which can be obtained from the derivatives of $\Phi_0(s,\infty)$ \cite{epidemic} is properly replicated as well when the heterogeneity in the number of recommendations is implemented (see Fig. \ref{casc}). Both results show how accurate the model is in predicting the reach of a viral marketing campaign by merely using its dynamic parameters. Moreover, since the values of $\lambda_1$ and $\overline{r}_{0},\overline{r}_{1}$ can be roughly estimated at the campaign early stages, we could have predicted its final reach at the very beginning.

\subsection{Time dynamics}

In the previous subsection we concentrate in the properties of the cascades in the asymptotic regime. Here we come back to the original equations for the dynamics of the nodes (\ref{eqf1}) and (\ref{eqf0}) to investigate its time dependence. Using on them that
\begin{equation}
i_{0,1} \equiv \langle I_{0,1}(t) \rangle = \left. \frac{\partial F_{0,1}(s,t)}{\partial s}\right|_{s=1}\end{equation}
we get
\begin{eqnarray}
i_0(t)&=&1-G_0(t) +  R_0\!\int_0^t \! \! dG_0(\tau) i_1(t-\tau)\\
i_1(t)&=&1-G_1(t) + R_1\!\int_0^t \! \! dG_1(\tau) i_1(t-\tau)
\end{eqnarray}
for the dynamics of the average number of infected participants.

Once again, the equation for $i_0(t)$ depends on the solution of the integral equation for $i_1(t)$. Actually, for $G_0(t) = G_1(t) = G(t)$ we could explicitly write $i_0(t) = [1-G(t)] + (R_0 / R_1) [i_1(t)-1+G(t)]$. However the solution for $i_1(t)$ is not known in general, although we can study its asymptotic behavior using Renewal Theory \cite{feller}. Such behavior strongly depends on the existence or not of the so called Malthusian Parameter $\alpha(\gamma,G)$ (\cite{branching} p.142), i.e. the real solution of the equation
\begin{equation}\label{malthusian}
\gamma \int_0^\infty e^{-\alpha t} dG(t) = 1
\end{equation}

If this parameter $\alpha = \alpha(\gamma,G)$ exists for $\gamma = R_1$ then $i_1(t)$ behaves asymptotically like
\begin{equation}\label{solmalthusian}
i_1(t) \sim C e^{\alpha t},\qquad C = \frac{R_1-1}{\alpha R_1^2 \int_0^\infty t e^{-\alpha t} dG(t)}
\end{equation}
for all values of $R_1$. Although $\alpha(\gamma,G)$ always exists above the ``tipping-point'' where $\gamma > 1$, there is a large class of distributions $G(t)$ for which $\alpha(\gamma,G)$ does not exist when $0< \gamma < 1$. This is the so called {\em sub-exponential} class which consists of all distribution functions $G(t)$ such that
\begin{equation}\label{asymp}
\lim_{t \to \infty} \frac{1-G^{*2}(t)}{1-G(t)}=2
\end{equation}
where $G^{*2}(t)$ is the twofold convolution of $G(t)$ \cite{athreya}. All those distributions have tails that decay slower than any exponential, that is, they are heavy-tailed distributions which is the best qualitative description of the sub-exponential class. Examples of $G(t)$ are power law (Pareto-like), stretched exponentials or log-normal distributions. For this class of distributions, the asymptotic behavior of $i_1(t)$ is given instead by the tail of the distribution
\begin{equation}\label{iasym}
i_1(t) \sim \frac{1-G(t)}{1-R_1}
\end{equation}

The asymptotic regime is reached for values of $t$ such that $1-G(t) \leq 1-R_1$ or, equivalently when $G(t) \geq R_1$. For the cascades size we get from equation (\ref{eqs1}) that
\begin{equation}
\langle S_1(t) \rangle = 1 + R_1\!\int_0^t \langle S_1(t-\tau)\rangle dG_1(\tau)
\end{equation}
whose asymptotic behavior, analyzed using Renewal Theory, gives
\begin{equation}
\langle S_1(t)\rangle \sim \left\{
\begin{array}{ll}
\langle S_1(\infty)\rangle - \frac{R_1}{1-R_1} i_1(t) & \quad \mathrm{if}\ R_1 < 1 \\
\frac{R_1}{R_1-1} i_1(t) & \quad \mathrm{if}\ R_1 > 1
\end{array}
\right.
\end{equation}

\section{Examples}\label{sec:examples}

In this section we illustrate two kinds of behavior that we can find in the time dynamics of the viral cascades. Specifically we consider the case in which $G(t)$ is super-exponential with two significant examples, the Poisson process and the Gamma process, and the case in which $G(t)$ is sub-exponential with application to the log-normal distribution found in section \ref{subexp}.

\subsection{Super-exponential processes}
When $G(t)$ is not sub-exponential the Malthusian parameter given by Eq. (\ref{malthusian}) always exists and the asymptotic solution is given by equation (\ref{solmalthusian}).

\medskip

\textbf{Poisson process:} Most of the literature assumes that $G(t)$ is the cdf of the exponential distribution for the response times. Thus, if $G_{0,1}(t) = 1-e^{-\rho_{0,1} t}$ equation (\ref{eqf1}) can be derived once to obtain
\begin{eqnarray}
\frac{\partial F_0(s,t)}{\partial t}&=&\rho_0 \{ f_0[F_1(s,t)]-F_0(s,t) \}\\
\frac{\partial F_1(s,t)}{\partial t}&=&\rho_1 \{ f_1[F_1(s,t)]-F_1(s,t) \}
\end{eqnarray}
and for the moments
\begin{eqnarray}\label{eqpoisson}
\frac{di_0}{dt} &=& \rho_0 [R_0 i_1(t) - i_0(t)]\\
\frac{di_1}{dt} &=& \rho_1[R_1-1]\, i_1(t)
\end{eqnarray}

The solution for the second equation with initial condition $i_1(0) = 1$ is $i_1(t) = e^{\alpha_1 t}$ with $\alpha_1 = \rho_1(R_1-1)$ and then
\begin{equation}
i_0(t) \sim \left\{
\begin{array}{ll}
\frac{R_0\rho_0}{\alpha_1+\rho_0} e^{\alpha_1 t} & \quad \mathrm{if}\ \alpha_1 \neq -\rho_0 \\
R_0\rho_0 t e^{-t\rho_0} &\quad \mathrm{if}\ \alpha_1 = -\rho_0
\end{array}
\right.
\end{equation}
where
\begin{equation}\label{malthusianpoissonian}
\alpha_1 = \rho_1(R_1-1) = \frac{R_1 -1}{\overline{\tau}_1}
\end{equation}
is the Malthusian parameter for $I_1(t)$. The resonant case $\alpha_1 = -\rho_0$ can only
happen below the ``tipping-point'' where $\alpha_1 <0$. Equations (\ref{eqpoisson}) are the linear growth Markovian models typically used to understand the dynamics of information spreading in social networks \cite{andersonmay}. In particular if the number of recommendations depends linearly with the substrate social network connectivity then $p_{1,r} \sim k p_{k} / \overline{k}$ and thus $R_1 = \lambda \overline{k^2}/ \overline{k}$ to recover the result by Pastor-Satorras and Vespignani \cite{satorras} that the Malthusian Parameter
\begin{equation}
\alpha_1 =  \lambda\frac{\overline{k^2}}{\overline{k}}-1\quad \Rightarrow\quad \lambda_c = \frac{\overline{k}}{\overline{k^2}}
\end{equation}

Thus, if the social connectivity has a distribution which is fat tailed then $\overline{k^2} \gg \overline{k}$ and $\lambda_c \simeq 0$. Moreover we recover the result of \cite{barthalemy} in which the Malthusian Parameter, in that case is $\alpha_1 \gg 1$, and leads to an exploding exponential that grows very fast in a short time.

\smallskip

The Poisson process is special, since $\alpha_1$ depends linearly on $R_1$. Thus the value of $R_1$ for social networks influences the total reach of the cascades but also the time dynamics. However this is not always the case, as we will see for other time processes. Besides, the Poissonian case tells us that the time dynamics of viral cascades is Markovian and that human dynamics can be described by differential equations like (\ref{eqpoisson}).

\smallskip

{\bf Gamma process:} In the case in which the distribution of response times is not given by an exponential, the behavior below the ``tipping-point'' is given by Eq. (\ref{solmalthusian}) for distributions $G(t)$ not in the sub-exponential class. Above the ``tipping-point'' the Malthusian parameter $\alpha$ always exists, but the relationship with $R_1$ can be highly non-linear. For example, in many applications it is found that the response time distribution $G(\tau)$ can be fitted to the cdf of the gamma distribution \cite{kalman,vazquezprl}, whose pdf is
\begin{equation}\label{pdfgamma}
P(\tau_1) = \tau_1^{k-1} \frac{e^{-\tau_1/\theta}}{\theta^k\Gamma(k)}
\end{equation}
where $\overline{\tau}_1 = k\theta$ and $\mathrm{Var}(\tau_1) = k\theta^2$. In fact, in \cite{vazquezprl} V\'azquez et al.\ found that the email response time is distributed as (\ref{pdfgamma}) with $k\simeq 0$ and $\theta \simeq 20$ days. On the other hand, the gamma distribution is used as simple model for the response time or lifetime since it can accommodate different functional behaviors: a delta function when $k \to \infty$ and $k\theta$ fixed, a power-law with exponential cutoff when $k<1$,  or the exponential case when $k= 1, 1/\theta = \rho$. For $k>0$ and $\theta < \infty$ the gamma distribution does not belong to the sub-exponential class. Thus the Malthusian parameter always exists and moreover it can be calculated exactly as
\begin{equation}\label{maltgamma}
\alpha_1 = \frac{R_1^{1/k}-1}{\theta}
\end{equation}

\begin{figure}
\centering
\includegraphics[width=0.45\textwidth,clip=]{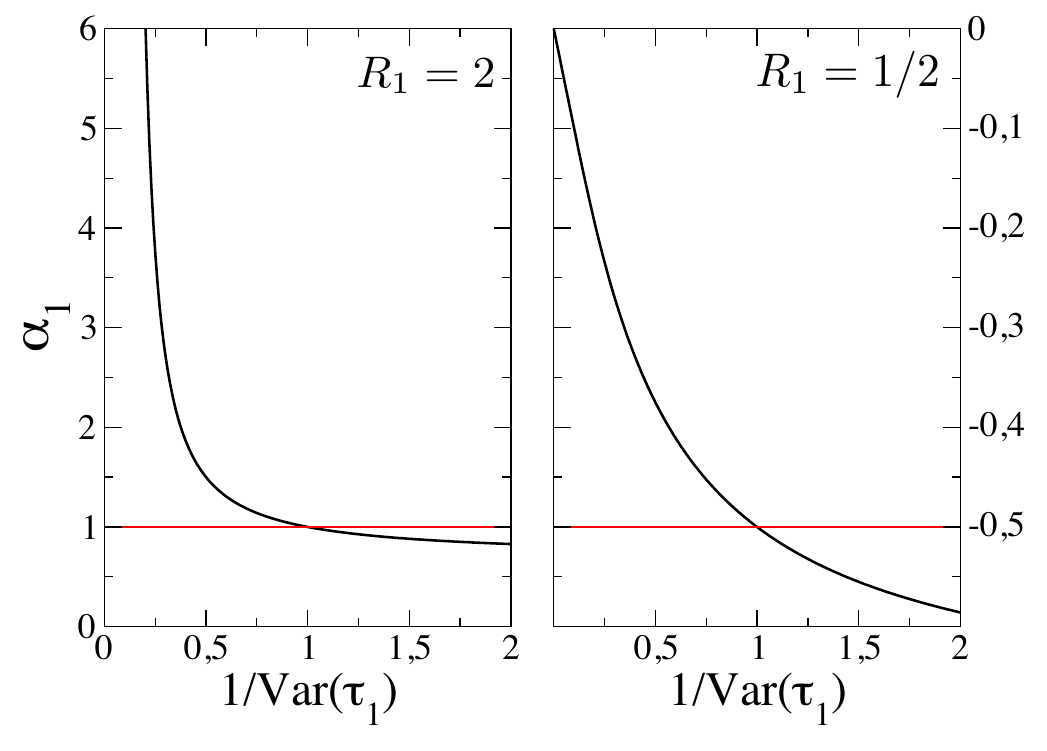}
\caption{(Color online) Malthusian parameter $\alpha_1$ for the gamma distribution of response times given by equation (\ref{maltgamma}) above (left, $\mu_1>1$) and below (right, $\mu_1 < 1$) the ``tipping-point''. The horizontal line is the malthusian parameter for the Poissonian approximation with the same average response time, i.e. $\rho = 1$. } \label{figmalth}
\end{figure}

This equations shows the non-trivial entanglement in the time dynamics of the recommendation process between the distribution of recommendations ($R_1$) and the response time distribution ($k,\theta$). In particular, it shows that the exponential growth depends not only on the mean response time $\overline\tau_1$ but also on the variance. To show this, we take the case $\overline{\tau}_1 = k\theta = 1$ fixed and we vary $k$ to control the variance. Figure \ref{figmalth} shows that above the ``tipping-point'' $\alpha_1$ diverges when $\mathrm{Var}(\tau_1)$ grows and thus propagation happens much more rapidly than in the case of the Poissonian approximation. The reason for it is that above the ``tipping-point'' the initial exponential growth of the infinite cascade is triggered by those people with response times below the mean, which in the case of long-tailed distributions are also more abundant than those with large response times. Below the ``tipping-point'', the contrary happens: since all cascades die out, their time dynamics is controlled by few nodes who, in the case of long-tailed distributions, can have large response times halting the branching process and slowing down the propagation of the information. In particular, Eq. (\ref{maltgamma}) recovers the result in \cite{vazquezprl} that with $k\simeq 0$ and {\em below} the ``tipping-point'' we get $\alpha_1 = -1/\theta$, i.e. the time scale is given by the cutoff in the distribution of response times.

\smallskip

However, it is important to note that even in this case, the asymptotic dynamics in the limit $t \to \infty$ is still given by the exponential decay in equation (\ref{solmalthusian}) which shows that although $\alpha_1$ now depends non-trivially on the moments of the $G(t)$ distribution we may describe the dynamics in terms of Markovian equations like (\ref{eqpoisson}) replacing $\alpha_1$ by its actual value.

\subsection{Sub-exponential process}\label{subexp}

In the case where $G(t)$ is sub-exponential the Malthusian Parameter does not exist below the ``tipping-point'' and the process asymptotic dynamics is given by the tail of the distribution $G(t)$ as Eq.\ (\ref{iasym}). In particular, this implies that we cannot describe the dynamics of viral cascades by Markovian approximations like the differential equations Eqs.\ (\ref{eqpoisson}) a sign for the strong non-Markovian character of the process in this situation, which corresponds to our empirical findings.

\begin{figure}
\centering
\includegraphics[width=0.48\textwidth,viewport= 1 7 310 220,clip=]{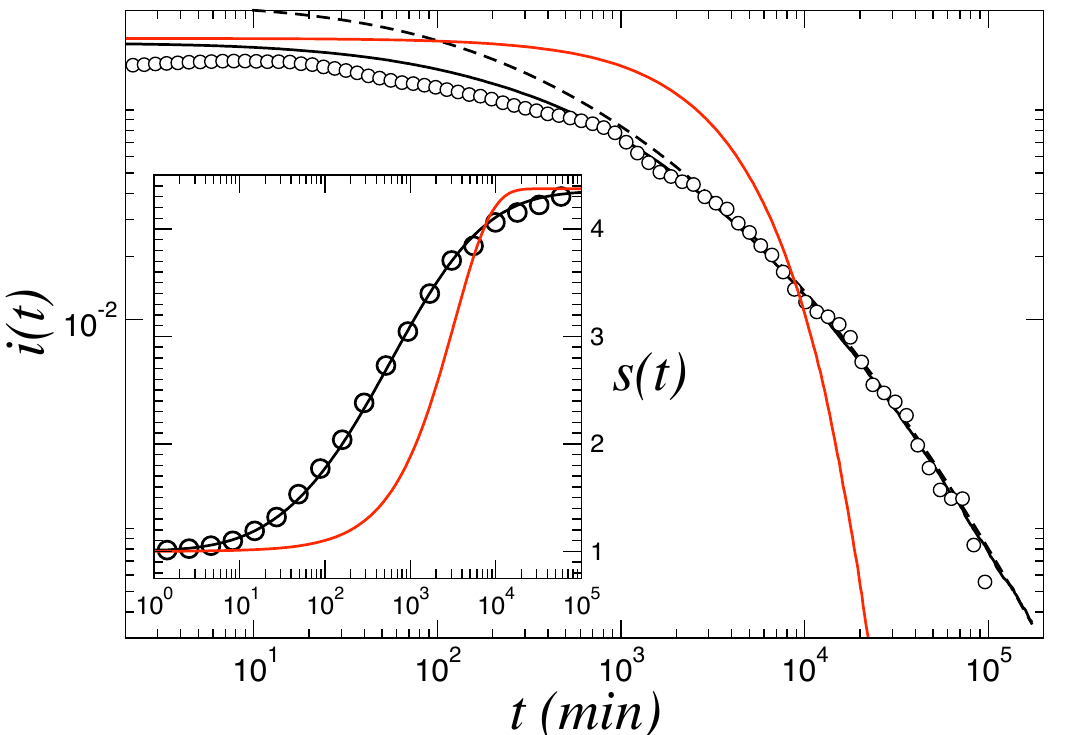}
\caption{(Color online) Distribution of the average fraction of new participants as a function of the cascades start time in our campaigns (circles) compared with Eq. (\ref{decay}) (black line), the prediction of the Bellman-Harris model with $P(t)$ the log-normal distribution of Eq. (\ref{lognor}) and with $P(t)$ exponential of the same mean (red). Dashed line is Eq. (\ref{decaylimit}), the asymptotic approximation of the Bellman-Harris model with $P(t)$ log-normal. Inset: Time dynamics of $S(t)$ the cascades average size (circles) accurately predicted by the model for $G(t)$ log-normal. In red predictions for $G(t)$ exponential.}\label{time}
\end{figure}

\smallskip

\textbf{Log-normal process:} We concentrate on the case where $G(t)$ is the cdf of the log-normal distribution which we found to be a good model for the response time in our campaigns. Specifically assuming its pdf is
\begin{equation}\label{lognor}
P(t)=\frac{1}{t \sigma_t \sqrt{2\pi}}e^{-(\ln t-\overline{\tau}_1)^2/(2\sigma_t^2)}
\end{equation} with mean $\overline{\tau}_1$ and variance $\sigma_t^2$, then Eq. (\ref{iasym}) tells us that
\begin{equation}\label{decay}
i_1(t)\sim\frac{1}{2(1-R_1)}\textrm{Erfc}\left(\frac{\ln t-\overline{\tau}_1}{\sigma_t\sqrt{2}}\right)
\end{equation}
where $\textrm{Erfc}(x)$ is the complementary error function. In the large $t$ limit we can replace $\textrm{Erfc}(x)$ in Eq. (\ref{decay}) by the first term of its asymptotic expansion to obtain
\begin{equation}\label{decaylimit}
i_1(t)\sim\frac{1}{(1-R_1)}\frac{\sigma_t}{\sqrt{2\pi}}\frac{\exp\left(-\frac{(\ln t-\overline{\tau}_1)^2}{2\sigma_t^2}\right)}{\ln t-\overline{\tau}_1}
\end{equation}
which indicates that the decay below the ``tipping-point'' is not exponential and also that it happens in a logarithmic and not in a linear time scale as shown in Fig. \ref{time}. This in turn implies that the information propagation prevails for much longer times than expected, as was shown in \cite{iribarren2009}, since the asymptotic dynamics in dying viral cascades can be dominated (and halted) by a single individual. However, above the ``tipping-point'' the malthusian parameter $\alpha_1$ always exists and can be calculated. In this case however, it can be very different from the Poissonian approximation given by Eq. (\ref{malthusianpoissonian}) since there is not an analytical solution in closed form for the Laplace transform of the log-normal distribution and equation \ref{malthusian} must be solved through numerical methods like the one proposed in \cite{shortle}. Finally, an important difference with the super-exponential process is that with sub-exponential cdf's of the response times Eq.\ (\ref{iasym}) shows an asymptotic dynamics for $i(t)$ that is always universally given by the cdf of the response (with a rescaling prefactor dependent on $R_1$). This could be used to measure $G(t)$ if no access to individual responses is possible. Note however than in the case of sub-exponential distributions, this is not possible since the Malthusian parameter in Eq. (\ref{malthusian}) depends highly non-trivially on both $G(t)$ and $R_1$.

\section{Conclusions}\label{sec:conclusions}

We closely tracked an invariable message propagation in an information diffusion process below the ``tipping-point'' (i.e. with $R_1<1$) driven by a real viral marketing mechanism run in several European markets. Our analysis of the data set of the resulting propagation that reached over 31,000 individuals, reveals the striking diffusion patterns that characterize the dynamics of information diffusion processes as being substantially different from the ones used in the epidemic models traditionally used to explain information propagation.

\smallskip

Those characteristic patterns affect both the structure of the  propagation paths and their dynamics. On the structural side, the viral propagation cascades are nearly pure trees almost completely devoid of closed loops or cycles and feature a very low clustering coefficient which is almost two orders of magnitude lower than the one typical of the email social networks upon which the viral propagation took place. Besides, the recommendations spreading activity of the campaigns active participants is very heterogeneous and its pdf is a long-tailed power-law which explains why most of the observed propagation was due to extraordinary events caused by \emph{super-spreading} individuals. On the other hand, the dynamics side of the propagation process shows that a majority of the spreading individuals become inactive right after sending their recommendations in what could be considered a ``birth-and-death'' process. Finally, the pdf of the forwarding time for the received recommendations is also a very heterogeneous long-tailed distribution, a log-normal in this case, and the spreaders forwarding time distribution and that of the number of recommendations they sent are independent and uncorrelated.

\smallskip

While there exist in the literature a number of studies about the static properties of viral information diffusion none of them explain the peculiar features discovered in the dynamics of our real campaigns. On the one hand most models concentrate only on the static asymptotic properties of the viral dynamics like Jurvetson's Viral Marketing model \cite{jurvetson}, the marketing percolation model of Goldenberg and Libai \cite{golden}, or the recommendation propagation model by Leskovec et al. \cite{huberman} which predicts a power-law with exponent $\gamma=-1$ for the distribution of the number of recommendations. On the other hand, numerous authors have studied the dynamic stochastic rumors \cite{yamir,daley,gani,rumors} using the Daley-Kendall (DK) or the Susceptible-Infective-Refractory (SIR) propagation models with Markovian differential equations, or the elaborate branching model of van der Lans et. al \cite{Lans2010}. However those models assume that the response time can be described by an exponential distribution which facilitates the theoretical analysis since Markovian and thus viral information diffusion can be explained by differential equations.

\smallskip

As we have found, this is not the case for our real experiments and we have described how to model the dynamics of information diffusion by means of the Bellman-Harris, which is the minimal framework to understand the non-Markovian spreading of information on social networks. This model generalizes the branching Galton-Watson scheme typically used both in information diffusion \cite{epidemic,golub,dwang,cebrian,Lans2010} and general percolation processes in social networks \cite{satorras,epidemic}. Our main result is that the information diffusion process object of this research shows a branching dynamics with some striking peculiarities that result a) from the human characteristic patterns when scheduling and prioritizing tasks, b) from the human decisions on how to select targets for the viral propagation, and, c) from the negligible influence of the substrate social network when the process runs below the ```tipping-point''. Thus, to explain all of them we propose a concise model that considers the large heterogeneity of human behavior but neglects the impact of the email social network underlying the diffusion process. The mathematical description of this approach is a non-Markovian, Bellman-Harris branching model with a sub-exponential (log-normal) distribution of the recommendations response time $G(t)$ like the one in Section \ref{subexp}, and two different power-law distributions for the number of referrals for the classes of \emph{Seed} and \emph{Viral} nodes, $p_{0,r}$ and $p_{1,r}$ respectively. Since $r_i$ and $\tau_i$ in our model are both iid random variables, the overall \emph{a priori} probability of transmission of the information between two individuals, the \emph{Transmissibility} $\lambda_1$, is the average over the distributions $p_{1,r}$ and $G_1(t)$ of the transmission probability between any two individuals \cite{epidemic}. Thus, per Newman \cite{siam}, our branching model is equivalent to uniform bond percolation on the same social network and several magnitudes of interest (cascades size distribution and ``tipping-point'') in the infinite time limit can be obtained by mapping it onto a bond percolation model.

\smallskip

Given the distributions $p_{0,r}$, $p_{1,r}$, $G_0(t)$ and $G_1(t)$, this model accurately predicts all the magnitudes of interest of the viral information or WOM diffusion processes: the dynamic parameters \emph{Transmissibility} and \emph{Fanout coefficient}, the cascades size distribution, its average and variance in the asymptotic limit, the \emph{cascades network} clustering coefficient, the message propagation ``tipping-point'' or the precise time dynamics in the asymptotic regime. Besides, it allows predictions for processes past, but close to, the ``tipping-point'' provided the substrate network of the propagation is large enough to avoid finite-size effects and maintain the assumption of its negligible influence. The accuracy of those predictions, which can be achieved early in the propagation process, make this model a valuable tool for managing information diffusion. Finally since most information transmission, sharing and searching in social networks has limited reach (thus happening below the ``tipping-point'') and given the fact that there seems to exists certain universality on both the heterogeneity in the number of actions \cite{barabasinature,telephone,blogs,blogs1,pitkow,tipping,sexual} and the sub-exponential character of human response times \cite{barabasinature,vazquez,amaralemail,iribarren2009,karsai,miritello}, our theoretical model is thus the most basic and general analytical tool to understand processes like rumor spreading, cooperation, opinion formation, cultural dynamics, diffusion of innovations, etc.

\appendix

\section{Clustering coefficients correlation}\label{appA}

\begin{figure}
\includegraphics[width=0.46\textwidth,viewport= 1 9 310 200,clip=]{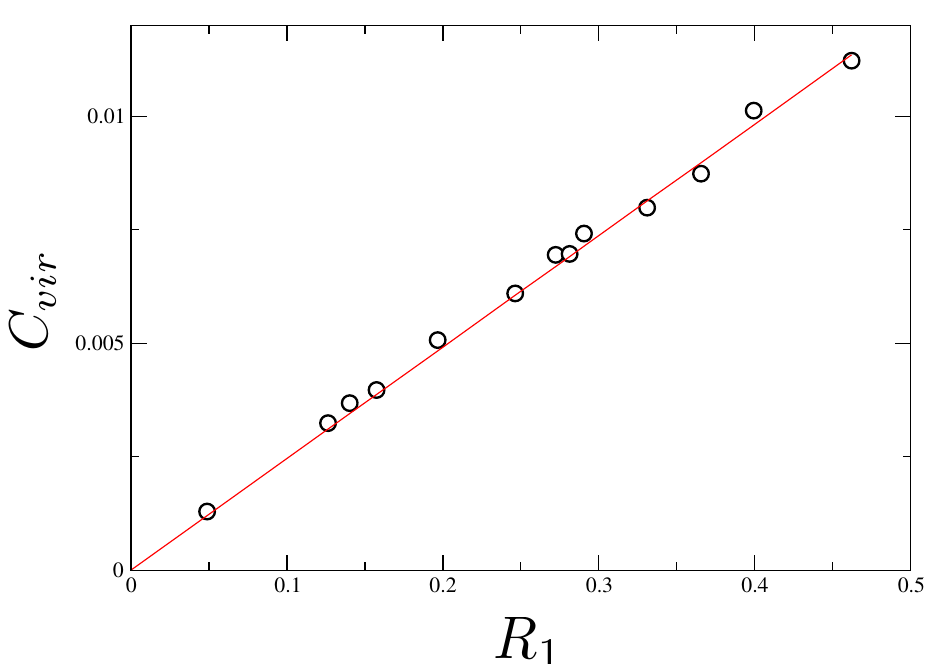}
\caption{Clustering coefficient $C_{cas}$ for the \emph{cascades network} obtained through simulations of the viral propagation model on a real email network (circles) with $C_{soc} = 0.22$ and $\langle \overline{k}_{nn}\rangle = 18.9$ compared with the linear relationship in Eq. (\ref{eq:cvir}).}\label{figclust}
\end{figure}

Assuming independence between the degree of a social network node and the number of messages it sends in a diffusion process, the undirected Clustering coefficients of the social network $C_{soc}$ and of the \emph{cascades network} $C_{cas}$ are correlated. Both are defined as \cite{siam}
\begin{equation}
C =\frac{3 \times \textrm{number of triangles in the network}}{\textrm{number of connected triples in the network}}\label{a1}
\end{equation}
where ``triple'' means a node with two edges running to an unordered pair of others. If connected, such pair forms a triangle. In a mean-field approximation they can be estimated as
\begin{eqnarray}
C_{soc} & = & \frac{3 \times \langle \textrm{triang}_{soc}\rangle}{\langle \textrm{tripl}_{soc}\rangle}\label{a2} \\
C_{cas} & = & \frac{3 \times \langle \textrm{triang}_{cas}\rangle}{\langle \textrm{tripl}_{cas}\rangle}\label{a3}
\end{eqnarray}
with $\langle \textrm{triang} \rangle$ and $\langle \textrm{tripl} \rangle$ being the average of triangles or triples by node in the social (\emph{soc}) or cascades (\emph{cas}) network. The probability of finding a triangle on a given node is the probability of it having a triple times the linking probability of its end nodes
\begin{equation}
P(\textrm{triang}) = P(\textrm{tripl}) \times P(\textrm{close})\label{a4}
\end{equation}
where $P(\textrm{close})$ is the existence probability of a link in the open side of the triad. Due to the independence of social links and recommendations, the average number of triangles and triples in the \emph{cascades network} results
\begin{eqnarray}
\langle \textrm{triang}_{cas} \rangle & = & P(\textrm{tripl}) \times P(\textrm{close}) \times \langle \textrm{triang}_{soc} \rangle\label{a5} \\
\langle \textrm{tripl}_{cas} \rangle & = & P(\textrm{tripl}) \times \langle \textrm{tripl}_{soc} \rangle\label{a6}
\end{eqnarray}
which replaced in \ref{a2}, \ref{a3} and combined with \ref{a4} yield
\begin{equation}
C_{cas} = P(\textrm{close}) \times C_{soc}\label{a7}
\end{equation}

Since nodes reached by a viral message become active with probability $\lambda$ and each resends it in average to $\overline{r}_1$ of its $\langle \overline{k}_{nn}\rangle-1$ nearest neighbors (excluding the ancestor node), the probability of closing the triple is
\begin{equation}
P(\textrm{close}) \sim \frac{2\lambda \overline{r}_1}{\langle \overline{k}_{nn}\rangle - 1}=\frac{2R_1}{\langle \overline{k}_{nn}\rangle - 1},  \qquad R_1\ll 1\label{a8}
\end{equation}
whose factor 2 stems from the fact that either of the nodes at the open end of a triple can send the message and close the triangle. Replacing $P(\textrm{close})$ in \ref{a7} recovers Eq. (\ref{eq:cvir}) which has been verified (even for $R_1\sim1$) through simulations on a university email network \cite{rovira} with $C_{soc} \sim 0.22$. Its correlation with the \emph{cascades network} Clustering coefficient as a function of $R_1$ is shown in Fig. \ref{figclust}. The low values of $C_{cas}$ explain why our model neglects the substrate network structure in the study of information propagation below the ``tipping-point''.


\begin{thebibliography}{99}

\bibitem{bruyn} A. De Bruyn and G. L. Lilien, Intern. J. of Research in Marketing \textbf{25}, 151, (2008).
\bibitem{buzzbuzz} R. Dye, Harvard Business Rev. \textbf{78}, 139 (2000).
\bibitem{Kumar2007} V. Kumar, J. A. Petersen, and R. P. Leone, Harvard Business Review R0710J, (2007).
\bibitem{Schmitt} P. Schmitt, B. Skiera, and C. Van den Bulte, Journal of Marketing \textbf{75}, 46, (2011).
\bibitem{barrat} A. Barrat, M. Barthelemy, and A. Vespignani, \emph{Dynamical processes on complex networks}, (Cambridge University Press, 2008).
\bibitem{castellano} C. Castellano, S. Fortunato and V. Loreto, Rev. Mod. Phys. \textbf{81}, 591, (2009).
\bibitem{wattsJCR} D. J. Watts, and P. S. Dodds, Journal of Consumer Research \textbf{34}, 441, (2007).
\bibitem{influence} D. Crandall, D. Cosley, D. Huttenlocher, J. Kleinberg, and S. Suri, KDD'08,ACM (2008)
\bibitem{iribarren2009}  J. L. Iribarren, and E. Moro, Phys. Rev. Lett. \textbf{103}, 038702, (2009).
\bibitem{aggregated} J. Phelps, R. Lewis, L. Mobilio, D. Perry, and N. Raman, Jour. of Advertising Research \textbf{44}, 333, (2005)
\bibitem{indirect1} C. Dellarocas, X. M. Zhang, and N. F. Awad, Jour. of Direct Marketing \textbf{21}, 23, (2007).
\bibitem{liben} D. Liben-Nowell, and J. Kleinberg, Proc. Natl. Acad. Sci. USA {\bf 105}, 4633, (2008).
\bibitem{epidemic} M. E. J. Newman, Phys. Rev. E \textbf{66}, 016128, (2002).
\bibitem{golub} B. Golub, and M. O. Jackson, Proc. Natl. Acad. Sci. USA \textbf{107}, 24, 10833 (2010).
\bibitem{dwang} D. Wang, Z. Wen, H. Tong, C. Y. Lin, C. Song, and A.-L. Barab\'asi, Proceedings of the 20th international conference on World Wide Web, \textbf{735}, (2011).
\bibitem{cebrian} G. Pickard, I. Rahwan, W. Pan, M. Cebrian, R. Crane, A. Madan, and A. Pentland, arXiv:1008.3172v1 [cs.CY], (2010).
\bibitem{yamir} Y. Moreno, M. Nekovee, and A. F. Pacheco, Phys. Rev. E \textbf{69}, 066130, (2004).
\bibitem{motion} C. A. Hidalgo, A. Castro, and C. Rodriguez-Sickert, New J. Phys. \textbf{8}, 52, (2006).
\bibitem{daley} D. J. Daley and D. G. Kendall, Nature \textbf{204}, 1118, (1964).
\bibitem{huberman} J. Leskovec, L. Adamic, and B. Huberman, ACM Transactions on the Web, \textbf{1}, (2007).
\bibitem{flow} F. Wu, B. A. Huberman, L.A. Adamic, and J. R. Tyler, Physica A \textbf{337}, 327 (2004).
\bibitem{barabasinature} A.-L. Barab\'asi, Nature {\bf 435}, 207, (2005).
\bibitem{telephone} W. Aiello, F. Chung, and L. Lu, Proc. of the 32nd Annual ACM Symposium of Theory of Computing, 171, ACM, New York, (2000).
\bibitem{blogs} D. Gruhl, R. Guha, D. Liben-Nowell, and A. Tomkins, Proc. of the 13th Int. Conf. on WWW, 491, (ACM, New York, 2004).
\bibitem{blogs1} J. Leskovec, M. McGlohon, and C. Faloutsos, Proc. of the SIAM Int. Conf. on Data Mining (SDM07, 2007).
\bibitem{pitkow} J. E. Pitkow, Proc. of the 7th WWW Conf. (WWW7, 1997).
\bibitem{tipping} M. Gladwell, \emph{The Tipping Point}, (Little, Brown and Company, New York, 2000).
\bibitem{sexual} F. Liljeros, C.R. Edling, L. A. N. Amaral, H. E. Stanley, and Y. Aberg, Nature \textbf{411}, 907 (2001).
\bibitem{andersonmay} R. M. Anderson, and R. May, \emph{Infectious diseases of humans: dynamics and control} (Oxford University Press, 1991).
\bibitem{vazquez} A. V\'azquez, J. G. Oliveira, Z. Dezs\"{o}, K. I. Goh, I. Kondor, and A.-L. Barab\'asi, Phys. Rev. E {\bf 73}, 036127 (2006).
\bibitem{amaralemail} D. B. Stouffer, R. D. Malmgren, and L. A. N. Amaral, Nature \textbf{235}, (2005).
\bibitem{karsai} M. Karsai, M. Kivel\"{a}, R. K. Pan, K. Kaski, J. Kert\'esz, A.-L. Barab\'asi, and J. Saram\"{a}ki, Phys. Rev. E \textbf{83}, 025102, (2011).
\bibitem{miritello} G. Miritello, E. Moro, and R. Lara, Phys. Rev. E {\bf 83}, 045102 (2011).
\bibitem{satorras} R. Pastor-Satorras, and A. Vespignani, Phys. Rev. Lett. {\bf 86}, 3200, (2001).
\bibitem{siam}  M. E. J. Newman, SIAM Review \textbf{45}, 167, (2003).
\bibitem{wattsviral} D. J. Watts, and J. Peretti, Harvard Business Rev., \textbf{F0705A}, (2007).
\bibitem{debruyn} A. De Bruyn, and G. L. Lillien, Int. Jour. of Research in Marketing 25, 143, (2008).
\bibitem{phonecalls} C. Kiss, A. Scholz, and M. Bichler, Proc. of the The 8th IEEE Int. Conf. on E-Commerce Technology and The 3rd IEEE Int. Conf. on Enterprise Computing, E-Commerce, and E-Services, (2006).
\bibitem{jurvetson} S. Jurvetson, and R. Draper, Netscape M-Files, (1997).
\bibitem{why} M. E. J. Newman, and J. Park, Phys. Rev. E {\bf68}, 036122, (2003).
\bibitem{ebelemail} H. Ebel, L.-I. Mielsch, and S. Bornholdt, Phys. Rev. E {\bf66}, 035103, (2002).
\bibitem{trust} C. Dellarocas, Management Science \textbf{49}, 1407, (2003).
\bibitem{feld} S. Feld, American Journal of Sociology \textbf{96}, 1464, (1991).
\bibitem{affinity} J. L. Iribarren, and E. Moro, Soc. Netw. \textbf{33}, 134, (2011).
\bibitem{ssdisease} J. O. Lloyd-Smith, S. J. Schreiber, P. E. Kopp, and  W. W.Getz, Nature {\bf 438}, 355, (2005).
\bibitem{bass} F. M. Bass, Management Science \textbf{15}, 215, (1969).
\bibitem{NM} M. E. J. Newman, S. Forrest, and J. Balthrop, Phys. Rev. E \textbf{66}, 035101 (R), (2002).
\bibitem{scalefree} A.-L. Barab\'{a}si and E. Bonabeau, Scientific American, \textbf{May}, 55, (2003).
\bibitem{melnik} S. Melnik, A. Hackett, M. A. Porter, P. J. Mucha, and J. P. Gleeson, Phys. Rev. E \textbf{83}, 036112, (2011).
\bibitem{branching} T. E. Harris, \emph{The Theory of Branching Processes}, (Springer Verlag, Berlin, 2002).
\bibitem{athreya} K. B. Athreya, and P. E. Ney, \emph{Branching Processes}, (Springer Verlag, Berlin, 1972).
\bibitem{Lans2010} R. van der Lans, G. van Bruggen, J. Eliashberg, and B. Wierenga, Marketing Science, \textbf{69}, 348, (2010).
\bibitem{dorogovtsev} S. N. Dorogovtsev, and J. F. F. Mendes, \emph{Evolution of Networks: From Biological Nets to the Internet and WWW}, (Oxford University Press, Oxford, 2003).
\bibitem{Karrer2010} B. Karrer, and M. E. J. Newman, Phys. Rev. E \textbf{82}, 016101, (2010).
\bibitem{feller} W. Feller, \emph{An Introduction to Probability Theory and Its Applications, Vol. I, Third Ed.}, (John Wiley \& Sons, New York, 1967).
\bibitem{barthalemy} M. Barthelemy, A. Barrat, R. Pastor-Satorras, and A. Vespignani, Jour. of Theoretical Biology \textbf{235}, 275, (2005).
\bibitem{kalman} Y.\ M.\ Kalman and S.\ Rafaeli. Proceedings of the 38th Annual Hawaii International Conference on System Sciences, 108 (2005).
\bibitem{vazquezprl} A.\ V\'azquez, B.\ R\'acz, A. Luk\'acs, and A.-L. Barab\'asi, Phys.\ Rev.\ Lett.\ {\bf 98}, 158702 (2007)
\bibitem{shortle} J. F. Shortle, M. J. Fischer, D. Gross, and D. M. B. Masi, Jour. of Probability and Statistical Science \textbf{1}, 15, (2003).
\bibitem{golden} J. Goldenberg, B. Libai, and S. Solomon, Physica A \textbf{284} 335, (2000).
\bibitem{gani} J. Gani, Environmental Modelling \& Software \textbf{15}, 721, (2000).
\bibitem{rumors} J. Zhou, Z. Liu, and B. Li, Physics Letters A \textbf{368}, 458, (2007).
\bibitem{rovira} R. Guimer\'{a}, L. Danon, A. D\'{i}az-Guilera, F. Giralt, and A. Arenas, Phys. Rev. E \textbf{68}, 065103, (2003).


\end{thebibliography}
\end{document}